\documentclass[a4paper,12pt]{article}

\pdfoutput=1 

\usepackage[hmargin=.71in,vmargin=1.1in]{geometry}
\usepackage{indentfirst}
\usepackage{amsfonts}
\usepackage{mathrsfs}
\usepackage{amsmath}
\usepackage{amssymb}
\usepackage{authblk}
\usepackage{cite}
\usepackage{xcolor}
\usepackage{mathtools}
\usepackage{tensor}
\usepackage{physics}
\usepackage{graphicx}
\usepackage{bm}
\usepackage{upgreek}
\usepackage{braket}
\usepackage{color,soul}
\usepackage{csquotes}
\usepackage{caption}
\usepackage{subcaption}
\usepackage{slashed}
\usepackage{mathtools}

\usepackage[bookmarksnumbered=true,bookmarksopen=true]{hyperref}
 \hypersetup{colorlinks,
             linkcolor=[rgb]{0,0.3,0.6}, 
             citecolor=[rgb]{0,0.3,0.6}, 
             urlcolor=[rgb]{0,0.3,0.6}}

\newcommand\mi{\mathrm{i}}
\newcommand\me{\mathrm{e}}

\newcommand\pp{\uppi}
\newcommand{\dif}{\mathrm{d}}

\begin{document}

\title{\Large\textbf{Phase diagrams of  quasinormal frequencies for Schwarzschild, Kerr,  and Taub-NUT black holes}}

\author[a]{Chen Lan\thanks{stlanchen@126.com}}
\author[b]{Meng-Hu Li\thanks{limhu@mail.nankai.edu.cn}}
\author[b]{Yan-Gang Miao\thanks{Corresponding author: miaoyg@nankai.edu.cn.}}

\affil[a]{\normalsize{\em Department of Physics, Yantai University, 30 Qingquan Road, Yantai 264005, China}}
\affil[b]{\normalsize{\em School of Physics, Nankai University, 94 Weijin Road, Tianjin 300071, China}}

\date{ }

\maketitle

\begin{abstract}
The Newman-Janis algorithm, which involves complex-coordinate transformations, establishes connections {\em between} static and spherically symmetric black holes {\em and} rotating and/or axially symmetric ones, such as between Schwarzschild black holes and Kerr black holes, and between Schwarzschild black holes and Taub-NUT black holes. However, the transformations in the two samples are based on different physical mechanisms. The former connection arises from the exponentiation of spin operators, while the latter from a duality operation. In this paper, we mainly investigate how the connections manifest in the dynamics of black holes. Specifically, we focus on studying the correlations of  quasinormal frequencies among Schwarzschild, Kerr, and Taub-NUT black holes. This analysis allows us to explore the physics of complex-coordinate transformations in the spectrum of quasinormal frequencies.
\end{abstract}

\tableofcontents

\section{Introduction}
\label{sec:intr}

Gravitational waves (GWs) were first predicted by Albert Einstein in his general theory of relativity in 1916. 
However, they were not observed until 2015  by the Laser Interferometer Gravitational-Wave Observatory (LIGO)~\cite{LIGOScientific:2016aoc,LIGOScientific:2016lio}. 
Since then, the detection of GWs has opened up~\cite{Neronov:2019uht,Meszaros:2019xej} a new era of multimessenger astronomy, 
where GW signals are combined with electromagnetic observations. 

The observation from the LIGO was the result of the merging of two black holes (BHs) in a binary system. 
Binary BHs  are pairs of BHs~\cite{Pretorius:2005gq} that orbit around each other. 
As they move closer together, they release energy in the form of GWs. 
This energy loss causes the BHs to spiral inward, eventually resulting in a cataclysmic merger. 
When the BHs merge, they create intense GWs that propagate outward through the universe. 
These waves carry crucial information about the astrophysical processes involved in the merger, as well as the properties of the BHs themselves.

Quasinormal modes (QNMs) or quasinormal frequencies (QNFs) are a fundamental concept~\cite{Konoplya:2011qq} in the study of GWs. 
This physical quantity describes a set of damping modes, where its 
real part determines the oscillation of GWs, 
while its imaginary part the damping rate that describes how quickly the oscillations decay over time. 
When two BHs merge, the emitted GWs change gradually from oscillation to exponential decay. 
These modes provide important information about the properties of the BHs. 
The analysis of QNFs in GW signals opens up new avenues for exploring the mysteries of the Universe and the fundamental nature of gravity.

The Newman-Janis algorithm (NJA)  is a {\em mathematical} method~\cite{Newman:1965tw} that exploits complex-coordinate transformations to convert a static and spherically symmetric BH solution to a rotating and/or axially symmetric one. 
In the study of regular black holes (RBHs)~\cite{Bambi:2013ufa,Toshmatov:2014nya,Azreg-Ainou:2014pra,Modesto:2010rv,Brahma:2020eos}, 
this algorithm is notable for two reasons. 
At first, it can generate rotating RBHs from a static seed, see the current review articles~\cite{Torres:2022twv,Lan:2023cvz} and the references therein. 
Secondly, it has the capability to modify or remove the curvature singularities of singular black holes (SBHs). 

As an example, the NJA can transform~\cite{Newman:1965tw} Schwarzschild BHs into Kerr BHs through the following transformations,
\begin{subequations}
\begin{equation}
\label{eq:transfer}
    u\to  u - \mi a \cos\theta,\qquad
    r\to r + \mi a \cos\theta, 
\end{equation}  
together with such complexifications,
\begin{equation} 
    \frac{1}{r} \to  \frac{\Re[r]}{\abs{r}^2},\qquad
    r^2\to \abs{r}^2.
\end{equation}
\end{subequations}
Here, $a$ denotes rotation parameter and $u$ ``time'' in the Eddington-Finkelstein coordinate. 
The underlying physical mechanism of the connection between Schwarzschild and Kerr BHs was established~\cite{Arkani-Hamed:2019ymq} through the {\em exponentiation} of spin operators, where a three-point  amplitude was considered in the minimal coupling of spinning particles and gravitons.

Now let us turn to the change in singularity.
The Kretschmann scalar of Schwarzschild BHs, which is a measure of curvatures, is proportional to $r^{-6}$ around $r=0$. 
After the complexification, $r^2$ becomes $\abs{r}^2$ and the radial coordinate takes a shift, $r\to r + \mi a \cos\theta$, 
and then the singular point $r=0$ changes into a singular ring described by $r^2 + a^2 \cos^2\theta=0$. 
In other words, the NJA alters the type of singular curvatures, from a point singularity to a ring singularity, when it is applied to Schwarzschild BHs.

As another example, by using the alternative transformations~\cite{Talbot:1969bpa}, 
\begin{subequations}
\label{eq:nja-nut}
\begin{equation}
    u\to  u - 2\mi N \ln \sin \theta,\qquad
    r\to  r-\mi N,\qquad
    M\to M-\mi N,
\end{equation}
where $N$ denotes a NUT charge, together with the corresponding complexifications, 
\begin{equation} 
    \frac{1}{r} \to  \frac{\Re[M \bar{r}]}{\abs{r}^2},\qquad
    r^2\to \abs{r}^2,
\end{equation}
\end{subequations}
where $\bar{r}$ denotes the complex conjugate of $r$, one can convert Schwarzschild BHs into Taub-NUT BHs. 
The relationship between Schwarzschild BHs and Taub-NUT BHs can be understood~\cite{Luna:2015paa} as a {\em duality} operation. 
In other words, it can be seen as a gravitational analog of electric-magnetic duality.
Moreover, the ``singularities'' are determined by the zeros of the algebraic equation, $r^2+N^2=0$, 
indicating that there is no curvature singularity along the real axis of $r$. 
This implies that the curvature singularity of Schwarzschild BHs has been removed. 
This phenomenon can also be seen in the Stokes portrait~\cite{Lan:2022qbb}, 
where the singularity is actually pushed onto the imaginary axis (nonphysical domain).

As demonstrated above, one can deduce rotating or axially symmetric BHs, Kerr or Taub-NUT, from the same static and spherically symmetric seed, Schwarzschild BHs, by using different transformations of the NJA. We are interested in {\em dynamical} differences hidden behind the different {\em mathematical} transformations because Kerr and Taub-NUT BHs are obviously distinct in some crucial properties, such as the singularity as mentioned above. In other words, we want to reveal the {\em physics} that is hidden behind {\em mathematics} (NJA transformations). Specifically, we mainly investigate how the different connections, {\em between} Schwarzschild {\em and} Kerr BHs and {\em between} Schwarzschild {\em and} Taub-NUT BHs, manifest in the QNFs, one of the significant features in dynamics of BHs.

The paper is arranged as follows. 
In Sec.~\ref{sec:curvature}, we analyze how the singularities of Kerr-Taub-NUT BHs change in the parameter space of $(a, N)$. 
We then present in Sec.~\ref{sec:light-ring/qnms} the analytical QNFs of Schwarzschild, Kerr, and Taub-NUT BHs through the light ring/QNMs correspondence. In order to acquire more accurate QNFs than the analytical ones, we need to perform numerical calculations as proceeded in the following three sections. 
In Sec.~\ref{sec:separation}, we discuss two types of test-field perturbations, scalar fields and spinor fields, with and without mass, where
we focus on the separation of variables. Further, we explore the spectrum of angular equations in Sec.~\ref{sec:angular}. 
We investigate the connections in the spectra of QNFs for Schwarzschild, Kerr, and Taub-NUT BHs in Sec.~\ref{sec:relation}. 
Finally, we present our conclusions in Sec.\ \ref{sec:conclusion}.
The Appendix  \ref{app:rec-coef} gives the coefficients of recursion formulas when we calculate the spectra of QNFs numerically by using Leaver's method.

\section{Kerr-Taub-NUT black holes and curvature invariants}
\label{sec:curvature}

To facilitate subsequent discussions, we combine the Kerr and Taub-NUT BHs into a single entity, referred to in literature as the Kerr-Taub-NUT spacetime~\cite{Miller:1973gao}. In the Boyer-Lindquist coordinates $(t,r,\theta,\phi)$, the metric of Kerr-Taub-NUT BHs can be expressed~\cite{Yang:2020iat} in the following form,
\begin{eqnarray}
\label{eq:Kerr-Taub-NUT}
\dif s^2&=&
    -\frac{\Delta}{\Sigma}
    \left[\dif t+(2 N \cos{\theta}-a \sin^2{\theta})\dif\phi\right]^2 +\frac{\Sigma}{\Delta}\dif r^2 \nonumber \\
& &+\frac{\sin^2{\theta}}{\Sigma}\left[-a \dif t+(r^2+a^2+N^2) \dif \phi\right]^2+\Sigma \dif\theta^2,
\end{eqnarray}
where $\Sigma$ and $\Delta$ are defined by
\begin{subequations}
\begin{equation}
\label{eq:metric-functions}
    \begin{split}
          \Sigma=\Sigma_1 \Sigma_2, \qquad \Sigma_1=r+\mi (a \cos{\theta}+N),\qquad
          \Sigma_2=r-\mi (a \cos{\theta}+N),
    \end{split}
\end{equation}
\begin{equation}
    \begin{split}
        \Delta=r^2-2 M r+ a^2- N^2.\label{defdelta}
    \end{split}
\end{equation}
\end{subequations}
Equation \eqref{eq:Kerr-Taub-NUT} describes Schwarzschild, Kerr, and Taub-NUT BHs, respectively, depending on the different regions of the parameter space  $(a, N)$.
\begin{itemize}
    \item If both $a$ and $N$ vanish, Eq.~\eqref{eq:Kerr-Taub-NUT} reduces to the metric of Schwarzschild BHs.
    \item If $a$ does not vanish but $N$ does, Eq.~\eqref{eq:Kerr-Taub-NUT} reduces to the metric of Kerr BHs.
    \item If $a$ vanishes but $N$ does not, Eq.~\eqref{eq:Kerr-Taub-NUT} reduces to the metric of Taub-NUT BHs.
\end{itemize}

Next, we turn to the curvature invariants of Kerr-Taub-NUT BHs, where they are composed of a complete set and referred to as Zakhary-Mcintosh invariants~\cite{Zakhary:2018odl}. This set contains seventeen elements and can be classified~\cite{Lan:2023cvz} into three groups; the Ricci type, solely constructed by Ricci tensors, the Weyl type, solely constructed by Weyl tensors, and the mixed type, constructed by both Ricci and Weyl tensors.

Because the Ricci tensor of Eq.\ \eqref{eq:Kerr-Taub-NUT} equals zero, $R_{\mu\nu}=0$, the curvature invariants derived by the contraction of Ricci tensors also equal zero. As a result, both the Ricci and mixed types are vanishing, and our calculations depend~\cite{Lan:2023cvz} only on the four elements in the Weyl type. The denominators of these four invariants are all proportional to the factor, $(N+a \cos \theta )^2+ r^2$, which gives the singularities as follows:
\begin{equation}
    r=0,\qquad N+a \cos \theta =0,
\end{equation}
or in the Cartesian coordinates  $(t,x,y,z)$ as follows~\cite{Misner:1973prb}: 
\begin{equation}
    x^2+y^2 = a^2 - N^2,\qquad
    z=0.
\end{equation}
Thus, we divide the singularities into three classes according to the parameter space $(a,N)$: 
\begin{itemize}
    \item If $a^2>N^2$, singular rings appear in the $x-y$ plane.
    \item If $a^2=N^2$, a singular point appears at the center.
    \item If $a^2<N^2$, no singularities appear.
\end{itemize}

We note that the singularities mentioned above are unrelated to the mass parameter. 
Furthermore, Kerr-Taub-NUT BHs manifest in two distinct phases in terms of the parameter space $(a,N)$ if the rotation parameter $a$ decreases from $a^2>N^2$ to $a^2<N^2$ for a fixed $N$ and simultaneously if the Kerr-Taub-NUT spacetime still exists. 
In one phase Kerr-Taub-NUT BHs contain singular rings in the case of $a^2>N^2$, while in the other phase Kerr-Taub-NUT BHs do not have curvature singularities in the case of $a^2<N^2$, where the two phases are separated by the configuration of Kerr-Taub-NUT BHs that possesses one singular point in the case of $a^2=N^2$.

\section{Analytical QNMs by the light ring/QNMs correspondence}
\label{sec:light-ring/qnms}

We provide the analytical QNMs of  Schwarzschild, Kerr, and Taub-NUT BHs using the light ring/QNMs correspondence~\cite{Cardoso:2008bp}, which connects the QNFs to circular null geodesics, known as photon spheres,
in the eikonal limit, 
\begin{equation}
\label{eq:qnfs}
    \omega = \Omega_c
l-\mi  \left(n+\frac{1}{2}\right) \lambda_c,
\end{equation}
where $\Omega_c$ denotes the angular velocity when a particle stays at an unstable null geodesic, $\lambda_c$ the Lyapunov exponent, $l$ the multipole number,
and $n$ the overtone number.

In order to determine the circular null geodesics of test particles in the Kerr-Taub-NUT spacetime, one calculates~\cite{Chakraborty:2013kza,Pradhan:2014zia} the effective potential of the radial equation of particles,
\begin{subequations}\label{EoMs}
\begin{equation}
\label{eq:eff-pot}
V_r = E^2 + \frac{2M r + 2N^2}{(r^2+N^2)^2}(a E-L)^2 + \frac{a^2 E^2-L^2}{r^2+N^2},
\end{equation}
the time-component equation with respect to the proper time,
\begin{equation}
\dot t = \frac{1}{\Delta} 
\left[
\left(r^2+N^2+a^2 +a^2\frac{2M r +2N^2}{r^2+N^2}\right) E
-\frac{a(2M r +2N^2)}{r^2+N^2}L
\right],\label{timecom}
\end{equation}
and the $\phi$ equation with respect to the proper time,
\begin{equation}
\dot\phi =\frac{1}{\Delta}\left[
\frac{r^2-2M r -N^2}{r^2+N^2}L
+\frac{2a(Mr +N^2)}{r^2+N^2} E
\right],\label{phicom}
\end{equation}
\end{subequations}
where $E$ and $L$ are the energy and angular momentum of test particles, respectively.
Further, one gives the radius of photon spheres by $V_r=0= V'_r$, 
\begin{equation}
\label{eq:photon-sphere}
r_c^3 - 3M r_c^2 -3N^2 r_c \pm 2a \sqrt{r_c(M r_c^2+2N^2 r_c - M N^2)} + M N^2 = 0.
\end{equation}
Thus, the 
angular frequency and Lyapunov exponent on the surface of photon spheres take the forms \cite{Dolan:2010wr}, 
\begin{equation}
\Omega_c = \abs{\dot\phi/\dot t}\Big|_{r=r_c},\qquad
\lambda_c=\left.\sqrt{\frac{V''_r}{2 \dot t^2}} \right|_{r=r_c},\label{omegalambda}
\end{equation}
from which one obtains that the angular frequency is exactly equal to the inverse of impact parameter $D_c$,
\begin{equation}
    \Omega_c =\frac{1}{\abs{D_c}},
    \qquad
    D_c^2 = a^2 +(r_c^2+N^2)
    \frac{3M r_c^2+4 N^2 r_c-M N^2}{M r_c^2+2N^2 r_c - M N^2}.\label{omegaimpact}
\end{equation}

Next, we shall compute the QNFs for the three BHs and compare their results.
However, prior to that, we would like to address a specific aspect of QNFs from the viewpoint of the NJA, i.e.,
we shall demonstrate that the QNFs of two BHs will exhibit a connection through the NJA if the two BHs are related by the NJA.

\begin{figure}[!ht]
     \centering
         \includegraphics[width=.65\textwidth]{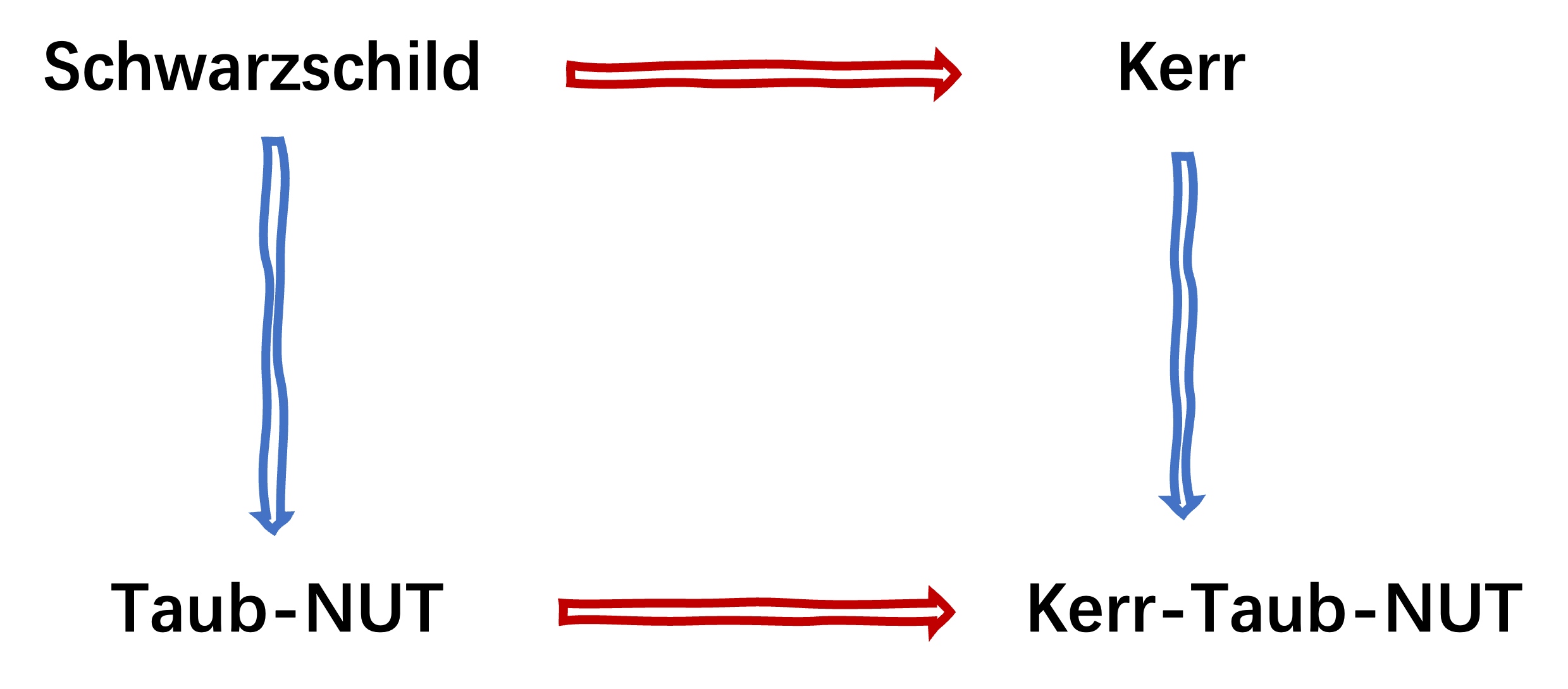}
      \captionsetup{width=.9\textwidth}
       \caption{Relationships of four models via  the NJA. 
       The blue arrows correspond to a simple relationship without mixing between coordinates and parameters, whereas
      the red arrows correspond to a complicated relationship with mixing between coordinates and parameters.}
        \label{fig:relation}
\end{figure}

It is usually considered that the NJA, as a {\em mathematical} method, converts the Schwarzschild metric to either the Kerr or Taub-NUT metric through distinct complex transformations. 
As a result, it is naturally anticipated that the {\em physics} will be interconnected through those complex transformations between Schwarzschild 
and Kerr BHs' QNFs or between Schwarzschild 
and Taub-NUT BHs' QNFs. To this end, we investigate the relationships of the four models (see Fig.~\ref{fig:relation}) connected by the NJA, where the Kerr-Taub-NUT BHs, as a single entity of Kerr and Taub-NUT BHs, are also contained in order to show a symmetric correlation.

To confirm the aforementioned assertion in Fig.~\ref{fig:relation} between the Schwarzschild and Taub-NUT metrics, 
or between the Kerr and Kerr-Taub-NUT metrics, see the blue arrows in Fig.~\ref{fig:relation}, a direct verification is possible owing to the fact that 
the complex transformations solely involve the radial coordinate and the parameter $M$, see Eq.~(\ref{eq:nja-nut}),
with no mixture of two coordinates, such as the radial and angular coordinates. 
In order to observe this, it is necessary to demonstrate that the three variables, $V_r$, $\dot t$, and $\dot \phi$, are interconnected between the Schwarzschild and Taub-NUT metrics 
or between the Kerr and Kerr-Taub-NUT metrics
through the consistent transformations Eq.~(\ref{eq:nja-nut}). 
The reason to make such a discussion is that the QNFs are associated solely with the three variables in the eikonal limit. Let us give the proof directly for the case from Schwarzschild to Taub-NUT BHs.\footnote{As to the case from Kerr to Kerr-Taub-NUT BHs, we can prove similarly but begin  from Eqs.~(\ref{eq:eff-pot})-(\ref{phicom}) with $N=0$.} 
At first, we reformulate the effective potential, see Eq.~(\ref{eq:eff-pot}) with $a=0$ and $N=0$ for Schwarzschild BHs, in the manner~\cite{Cardoso:2008bp}, 

\begin{equation}
V_{\rm Sch} = E^2 + \frac{2M r }{(r^2)^2}L^2 - \frac{L^2}{r^2}.
\end{equation}
Then following the complex transformations, Eq.~\eqref{eq:nja-nut},  we transform $V_{\rm Sch}$ into the form, 
\begin{equation}
\widetilde{V} = E^2 + \frac{2 \Re[M \bar{r}] }{(\abs{r}^2)^2}L^2 - \frac{L^2}{\abs{r}^2},
\end{equation}
which is just the effective potential of the Taub-NUT spacetime. 
Similarly, starting from $\dot t$ and $\dot \phi$ for Schwarzschild BHs, see Eqs.~(\ref{timecom}) and (\ref{phicom}) together with $a=0$ and $N=0$,
\begin{equation}
	\dot t_{\rm Sch} = \frac{1}{\Delta_{\rm Sch}} 
	r^2 E,
\end{equation}
and
\begin{equation}
	\dot\phi_{\rm Sch}=
	\frac{L}{r^2},
\end{equation} 
we derive their Taub-NUT forms by using Eq.~\eqref{eq:nja-nut},
\begin{equation}
\dot{\widetilde{t}} = \frac{1}{\widetilde{\Delta}} 
\abs{r}^2  E,
\end{equation}
and
\begin{equation}
\dot{\widetilde{\phi}}=
\frac{L}{\abs{r}^2},
\end{equation}
where $\Delta_{\rm Sch}$, see Eq.~(\ref{defdelta}) together with $a=0$ and $N=0$, and its transformed form read
\begin{equation}
	\Delta_{\rm Sch}=r^2-2 M r,
\end{equation}
and
\begin{equation}
	\widetilde{\Delta}=\abs{r}^2-2 \Re[M \bar r].
\end{equation}
The QNFs of Taub-NUT BHs can be understood as a distortion of the QNFs of Schwarzschild BHs under the complex transformations depicted by Eq.~\eqref{eq:nja-nut}.

However, the relationships between Schwarzschild and Kerr BHs, or between Taub-NUT and Kerr-Taub-NUT BHs in  $V_r$, $\dot t$, and $\dot \phi$, see the red arrows in Fig.~\ref{fig:relation},
are more intricate owing to the combination of radial and angular coordinates in the complex transformations, even for geodesics in the equatorial plane. 
Nonetheless, the indications of their relevance to QNFs can be observed. 
In the complex transformations from Schwarzschild to Kerr BHs, see Eq.~(\ref{eq:transfer}),
the additional introduction of nondiagonal metric components is needed, which results in the mixtures between energy $E$ and angular momentum $L$, 
and between $\dot t$ and $\dot\phi$, as shown in Eqs.~(67) and (68) of Ref.~\cite{Cardoso:2008bp}. 
These mixtures lead to complexity in the structure of QNFs, which is commonly referred to as {\em Zeeman splittings} in literature, e.g., Ref.~\cite{Berti:2004md}, and will be elaborated upon below.

In contrast with the case between Schwarzschild and Kerr BHs, a self-dual Taub-NUT BH with mass $M$ equal to $\pm N$ can be transformed~\cite{Crawley:2021auj,Crawley:2023brz,Guevara:2023wlr} into a self-dual Kerr-Taub-NUT BH through complex rotations of variables and parameters when the Kleinian signature $(--++)$ or even the Euclidean signature $(++++)$ is considered. 
This may suggest that the splittings of the spectrum of QNFs caused by the presence of rotation parameter $a$ are likely attributed to the complex transformations' multiple values.

\subsection{Schwarzschild black holes}

For Schwarzschild BHs, $a=N=0$, the equation of photon spheres has only one root outside the horizon, i.e., 
we derive the radius of horizons and the radius of photon spheres from Eq.~(\ref{eq:photon-sphere}), 
\begin{equation}
    r^{\rm Sch}_{\rm H} = 2M,\qquad
     r^{\rm Sch}_c =3 M,\label{radii}
\end{equation}
respectively, and then the impact parameter using Eq.~(\ref{omegaimpact}),
\begin{equation}
    D^{\rm Sch}_c = 3\sqrt{3}M.
\end{equation}
As a result, we conclude that the angular frequency equals the Lyapunov exponent by considering Eqs.~(\ref{EoMs}), (\ref{omegalambda}), and (\ref{radii}),
\begin{equation}
   \Omega_c^{\rm Sch}  = \lambda^{\rm Sch}_c= \frac{1}{3\sqrt{3} M},\label{schomelam}
\end{equation}
which contains all the information of QNFs in the eikonal limit based on the light ring/QNMs correspondence Eq.\ \eqref{eq:qnfs}.

\subsection{Kerr black holes}

In the case of Kerr BHs, the existence of a horizon depends on the condition that $|a|$ is less than $M$. 
This condition gives the horizon radius of Kerr BHs,
\begin{equation}
r^{\rm Ker} _{\rm H}=M+\sqrt{M^2-a^2}.
\end{equation}
Furthermore, the radii of photon spheres for Kerr BHs take~\cite{Dolan:2010wr,Yang:2012he,Fransen:2023eqj}
three values depending on the types of orbits:  corotating, counterrotating, and polar,
\begin{subequations}
\begin{equation}
\label{eq:equatorial-ring}
r^{\rm Ker} _{\pm} = 2M \left[
1+\cos\left(
\frac{2}{3}\cos^{-1}\left(\mp\frac{a}{M}\right)
\right)
\right],
\end{equation}
\begin{equation}
r^{\rm Ker} _{o} = M +2\sqrt{M^2-a^2}\cos\left[
\frac{1}{3}\cos^{-1}\left(
\frac{M(M^2-a^2)}{(M^2-a^2/3)^{3/2}}
\right)
\right],
\end{equation}
\end{subequations}
where the subscripts $\pm$ and $o$ represent corotating, counterrotating, and polar orbits, respectively. 
It is important to highlight that the radii of photon spheres for corotating and polar orbits are smaller than $3M$ (Schwarzschild BHs), 
\begin{equation}
0<r^{\rm Ker} _{+}<3M,\qquad
0<r^{\rm Ker} _{o}<3M,
\end{equation}
while the radius for a counterrotating orbit is larger than $3M$,
\begin{equation}
r^{\rm Ker} _{-}>3M.
\end{equation}
Additionally, a mirror symmetry can be observed between corotating and counterrotating orbits with respect to the rotation parameter $a$,
\begin{equation}
r^{\rm Ker} _{+}(a)  = r^{\rm Ker} _{-}(-a).
\end{equation}
This symmetry is also applicable to the angular velocity and Lyapunov exponent, as shown below.

Substituting $V_r$, $\dot t$, and $\dot \phi$ of Kerr black holes into Eq.~(\ref{omegalambda}), we obtain the angular velocities for the three obits,
\begin{equation}
\Omega^{\rm Ker}_\pm =\frac{1}{\abs{D_\pm}}= \frac{M^{1/2}}{(r_\pm^{\rm Ker})^{3/2}\pm a M^{1/2}},\qquad
\Omega^{\rm Ker}_o = \frac{\pp }{2 \sqrt{D_0^2-a^2} E\left(\frac{\mi a}{\sqrt{D_0^2-a^2}}\right)},
\end{equation}
where
$D_\pm=\pm 3\sqrt{M r^{\rm Ker}_\pm}-a$ are impact parameters for corotating and counterrotating orbits, respectively,
$D_o$ is the impact parameter for the polar orbit,
\begin{equation}
	D_o^2 =\frac{(3r_o^2-a^2)(r_o^2+a^2)}{r_o^2-a^2},
\end{equation}
and $E(k)=\int^{\pp/2}_0 \sqrt{1-k^2\sin^2\theta} \dif \theta$ is the complete elliptic integral of the second kind. Moreover, we derive
the Lyapunov exponents for the three orbits,
\begin{subequations}
\begin{equation}
\lambda^{\rm Ker}_\pm =\Omega^{\rm Ker}_\pm \frac{1-2 a /D_\pm}{\sqrt{1-a^2 /D_\pm^2}},
\end{equation}
\begin{equation}
\lambda^{\rm Ker}_o=\frac{r^{\rm Ker}_o}{D_0^2-a^2}\frac{K(x_o/\sqrt{1+x_o^2})}{\sqrt{1+x_o^2}E(\mi x_o)}\left[3-\frac{a^2(D_o^2-a^2)}{(r_o^{\rm Ker})^4}\right]^{1/2},
\end{equation}
where $x_o=a/(D_o^2-a^2)$ and $K(k)=\int^{\pp/2}_0 \frac{\dif\theta}{\sqrt{1-k^2 \sin^2\theta}}$ is 
the complete elliptic integral of the first kind. 
\end{subequations}

Since $D_+(-a)=-D_-(a)$, the angular velocities and Lyapunov exponents possess the mirror symmetry under the transformation of $a\to-a$, 
\begin{equation}
\Omega^{\rm Ker}_+(a)=\Omega^{\rm Ker}_-(-a),\qquad
\lambda^{\rm Ker}_+(a)=\lambda^{\rm Ker}_-(-a).
\end{equation}
This symmetry implies~\cite{Onozawa:1996ux,Berti:2009kk,Konoplya:2017tvu} that the QNFs for corotating and 
counterrotating orbits are not independent.

\subsection{Taub-NUT black holes}

The horizon of Taub-NUT BHs is not less than that of Schwarzschild BHs,
\begin{equation}
    r_{\rm H}^{\rm NUT}= 
    M+\sqrt{M^2+N^2}.
\end{equation}
The radius of photon spheres can be cast in the form similar to that of Kerr BHs,
but with a different parameter $z$ owing to the substitution of $a=0$ into Eq.~\eqref{eq:photon-sphere}, 
\begin{equation}
r_c^{\rm NUT}=
M+
z \sqrt{M^2+N^2},\qquad
z=2 \cos \left(\frac{1}{3} \tan^{-1} \frac{N}{M}\right),\label{radiustaub}
\end{equation}
where $z$ is a monotonic decreasing function of $N/M$ and has the following limits,
\begin{equation}
    \lim_{N/M\to0} z = 2,\qquad
    \lim_{N/M\to\infty} z = \sqrt{3}>1,
\end{equation}
which, like the case of Kerr BHs, shows that the photon sphere is outside the horizon. Substituting $a=0$ into Eq.~(\ref{omegaimpact}), we obtain the angular velocity and the impact parameter,
\begin{equation}
    \Omega_c^{\rm NUT}=\frac{1}{D_c^{\rm NUT}}= \sqrt{\frac{M \left(r_c^2-N^2\right)+2 N^2 r_c}{\left(N^2+r_c^2\right) \left(-M N^2+3 M r_c^2+4 N^2 r_c\right)}},
\end{equation}
and using Eqs.~(\ref{EoMs}), (\ref{omegalambda}), and (\ref{radiustaub}), we can derive the Lyapunov exponent from the following ratio,
\begin{equation}
    \frac{\lambda_c^{\rm NUT}}{\Omega_c^{\rm NUT}} = 
    \frac{ \sqrt{ 3\left[2 N^2 r_c (3 r_c-2 M)+r_c^3 (4M-r_c)-N^4\right]}}{ N^2+r_c^2}.\label{reimtaub}
\end{equation}

\subsection{Analysis of quasinormal frequencies}
\label{sec:num-qnf}

We give two comments about QNFs depicted by Eq.~\eqref{eq:qnfs}.
The first is that Eq.~\eqref{eq:qnfs} is applicable only under the eikonal limit, 
where the multipole number (angular momentum) $l$ is much larger than one and then its contribution is much larger than that of spins. 
Consequently, the QNFs do not encompass any spin characteristics.
The second comment is that the real part of QNFs is affected by the multipole number $l$ but not by the overtone number $n$, 
and the imaginary part is affected by the overtone number $n$ but not by the multipole number $l$. 
This implies that the real part displays a positive correlation with the multipole number $l$, and the imaginary part does with the overtone number $n$.

Now let us analyze the asymptotic behaviors of $\Omega_c$ and $\lambda_c$.
When the rotation parameter $a\to 0$ for Kerr BHs and the NUT charge $N\to0$ for Taub-NUT BHs, the leading terms of $\Omega_c$ and $\lambda_c$ for the two BHs must be
consistent with the angular velocity and Lyapunov exponent of Schwarzschild BHs,
\begin{subequations}
\label{keromelam}
\begin{equation}
    \Omega_\pm^{\rm Ker}\sim\frac{1}{3 \sqrt{3} M}\pm\frac{2 a}{27 M^2}+O\left(a^2\right),
\end{equation} 
\begin{equation}
    \lambda_\pm^{\rm Ker} \sim \frac{1}{3 \sqrt{3} M}-\frac{2 a^2}{81 \sqrt{3} M^3}+O\left(a^3\right),
\end{equation}
\end{subequations}
\begin{subequations}
\begin{equation}
\Omega^{\rm Ker}_o\sim\frac{1}{3 \sqrt{3} M}+\frac{7 a^2}{324 \sqrt{3} M^3}+O\left(a^3\right),
\end{equation}
\begin{equation}
\lambda^{\rm Ker}_o\sim\frac{1}{3 \sqrt{3} M}-\frac{a^2}{54 \sqrt{3} M^3}+O\left(a^3\right),
\end{equation}
\end{subequations}
and
\begin{subequations}
\label{nutomelam}
\begin{equation}
    \Omega_c^{\rm NUT}\sim
    \frac{1}{3 \sqrt{3} M}
    -\frac{5 N^2}{54 \sqrt{3} M^3}+O\left(N^3\right),
\end{equation} 
\begin{equation}
    \lambda_c^{\rm NUT} \sim 
    \frac{1}{3 \sqrt{3} M}
    -\frac{11 N^2}{162 \sqrt{3} M^3}+O\left(N^3\right).
\end{equation}
\end{subequations}
This is not difficult for us to understand from physics since both Kerr and Taub-NUT BHs reduce to Schwarzschild BHs when $a\to0$ and $N\to 0$, respectively.

In accordance with Eqs.~(\ref{schomelam}), (\ref{keromelam}), and (\ref{nutomelam}), we compare angular velocities and Lyapunov exponents between Schwarzschild and Kerr BHs, between Schwarzschild and Taub-NUT BHs, and between Kerr and Taub-NUT BHs, respectively,
\begin{equation}
\label{eq:slope-ks}
\Omega_{+,o}^{\rm Ker}/\Omega_c^{\rm Sch}>1,\qquad
   \lambda_{+, o}^{\rm Ker}/\lambda_c^{\rm Sch}<1,
\end{equation}
\begin{equation}
\Omega_{-}^{\rm Ker}/\Omega_c^{\rm Sch}<1,\qquad
   \lambda_{-}^{\rm Ker}/\lambda_c^{\rm Sch}<1,
\end{equation}
\begin{equation}
\label{eq:slope-ns}
   \Omega_c^{\rm NUT}/\Omega_c^{\rm Sch}<1,\qquad
    \lambda_c^{\rm NUT}/\lambda_c^{\rm Sch}<1,
\end{equation}
\begin{equation}
\label{eq:slope-kn}
 \Omega_c^{\rm NUT}/\Omega_{+,o}^{\rm Ker}<1,\qquad
    \lambda_c^{\rm NUT}/\lambda_{+,o}^{\rm Ker}<1,
\end{equation}
\begin{equation}
\Omega_c^{\rm NUT}/\Omega_{-}^{\rm Ker}>1,\qquad
    \lambda_c^{\rm NUT}/\lambda_{-}^{\rm Ker}<1,
\end{equation}
where we have applied the limit of $a=N\to0$ in the comparison of asymptotic behaviors  between Kerr and Taub-NUT BHs.
Based on the above inequalities, 
we observe that
the rotation parameter $a$ is associated with an increase in the oscillation frequency for corotating and polar orbits ($\Omega_{+,o}^{\rm Ker}/\Omega_c^{\rm Sch}>1$), 
whereas the NUT charge $N$ is linked to a decrease in the oscillation frequency ($\Omega_c^{\rm NUT}/\Omega_c^{\rm Sch}<1$ and $\Omega_c^{\rm NUT}/\Omega_{+,o}^{\rm Ker}<1$). Moreover, we notice that
both the rotation parameter $a$ and the NUT charge $N$ result in a weakening decay ($\lambda_{\pm,o}^{\rm Ker}/\lambda_c^{\rm Sch}<1$, $\lambda_c^{\rm NUT}/\lambda_c^{\rm Sch}<1$, and $\lambda_c^{\rm NUT}/\lambda_{\pm,o}^{\rm Ker}<1$). 
Therefore, we may refer Kerr BHs as a counterpart of Schwarzschild BHs with an increasing frequency owing to $\Omega_{+,o}^{\rm Ker}-\Omega_c^{\rm Sch}>0$, while  Taub-NUT BHs as a counterpart of Schwarzschild BHs with a decreasing frequency owing to $\Omega_c^{\rm NUT}-\Omega_c^{\rm Sch}<0$.

Opposite to the limit of $a\to 0$ for Kerr BHs, 
we now consider the limit\footnote{The rotation parameter $a$ cannot be greater than the mass of Kerr BHs, otherwise there are no horizons.} of $a\to M$, under which $\Omega_+^{\rm Ker}$ goes to a constant $1/(2M)$, and $\lambda_+^{\rm Ker}$ vanishing,
\begin{subequations}
\begin{equation}
    \Omega_+^{\rm Ker}\sim
    \frac{1}{2 M}+
    \frac{\sqrt{3\left(M-a\right)}}{2\sqrt{2} M^{3/2}}+O\left((a-M)^1\right),
\end{equation} 
\begin{equation}
    \lambda_+^{\rm Ker} \sim 
    -\frac{ \sqrt{M-a}}{\sqrt{2} M^{3/2}}
    +O\left((a-M)^1\right),
\end{equation}
which implies that $\Omega_+^{\rm Ker}/\Omega^{\rm Sch}>1$ 
and $\lambda_+^{\rm Ker}$ vanishes as $a$ approaches the limit value $M$.
In other words, when $a$ approaches $M$, the corotating orbit of Kerr BHs reaches a stable state without any decay. 
\end{subequations}
For the counterrotating orbit, we obtain
\begin{subequations}
\begin{equation}
    \Omega_-^{\rm Ker}\sim
   \frac{1}{7 M}+\frac{5 (M-a)}{147 M^2}+O\left((a-M)^2\right),
\end{equation} 
\begin{equation}
    \lambda_-^{\rm Ker} \sim 
   \frac{3 \sqrt{3}}{28 M} +\frac{5 (M-a)}{294 \sqrt{3} M^2}+O\left((a-M)^2\right),
\end{equation}
\end{subequations}
which implies that the counterrotating orbit is an unstable state because of a  finite value of $\lambda_-^{\rm Ker}$  and that the rotation parameter makes a decreasing effect to the angular velocity ($\Omega_-^{\rm Ker}/\Omega^{\rm Sch}<1$) as $a$ approaches $M$.
For the polar orbit, we derive
\begin{subequations}
\begin{equation}
    \Omega_o^{\rm Ker}\sim
   \frac{0.20937}{M}-\frac{0.0489321 (M-a)}{M^2}+O\left((a-M)^2\right),
\end{equation} 
\begin{equation}
    \lambda_o^{\rm Ker} \sim 
   \frac{0.165616}{M}+\frac{0.130548 (M-a)}{M^2}+O\left((a-M)^2\right).
\end{equation}
\end{subequations}
which implies that the polar orbit is an unstable state because of a  finite value of $\lambda_o^{\rm Ker}$  and that the rotation parameter makes an increasing effect to the angular velocity ($\Omega_o^{\rm Ker}/\Omega^{\rm Sch}>1$) as $a$ approaches $M$.
For Taub-NUT BHs, both $\Omega_c^{\rm NUT}$ and $\lambda_c^{\rm NUT}$ vanish when the NUT charge goes to infinity instead of zero,
\begin{subequations}
\begin{equation}
    \Omega_c^{\rm NUT}\sim
    \frac{1}{2 \sqrt{2} N}+O\left(N^{-2}\right),
\end{equation} 
\begin{equation}
    \lambda_c^{\rm NUT} \sim \frac{\sqrt{3}}{4 N}
    +O\left(N^{-2}\right).
\end{equation}
\end{subequations}

\begin{figure}[!ht]
     \centering
         \includegraphics[width=.65\textwidth]{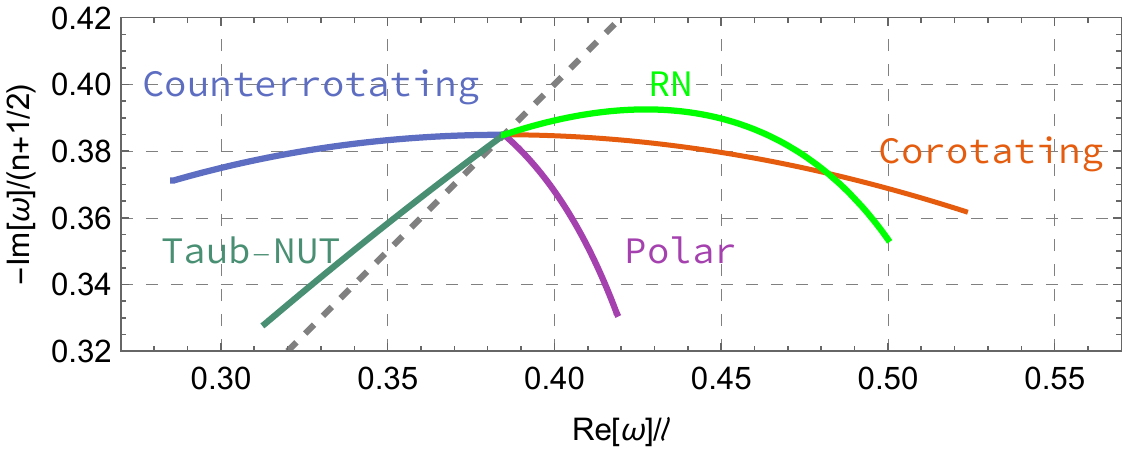}
      \captionsetup{width=.9\textwidth}
       \caption{The Zeeman splitting of QNFs, where the horizontal axis denotes Re($\omega)/l$, i.e., the angular velocity $\Omega_c$, and the vertical axis stands for $-{\rm Im}(\omega)/(n+1/2)$, i.e., the Lyapunov exponent $\lambda_c$. The dashed gray line represents the case of Schwarzschild BHs, where the QNFs change with respect to the mass $M$. The red, blue, and purple curves give the QNFs of Kerr BHs for corotating, counterrotating, and polar obits, respectively, where $M$ takes $1/2$, which yields the same results as those of the dimensionless treatment if $2M$ is chosen to be a normalization factor. The dark green curve denotes the QNFs of Taub-NUT BHs that change with respect to the NUT charge $N$, where $M$ also takes $1/2$. The light green curve shows the case of RN BHs, i.e., the QNFs vary with the charge for a fixed  $M=1/2$.}
        \label{fig:zeeman-splitting}
\end{figure}

Combining the above properties of the QNFs for Schwarzschild, Kerr, and Taub-NUT BHs,
we depict the QNFs in Fig.~\ref{fig:zeeman-splitting}, where the QNFs of
Reissner-Nordstr\"om (RN) BHs are attached as a comparison. This diagram illustrates how the QNFs vary with the  parameters, such as the mass $M$ for Schwarzschild BHs, the rotation parameter $a$ and mass $M$ for Kerr BHs, and the NUT charge $N$ and mass $M$ for Taub-NUT BHs, and with the parameters --- electric charge $Q$ and mass $M$ for RN BHs.

The QNFs of Schwarzschild BHs, see Eq.~\eqref{schomelam}, 
which change only with the mass $M$, divide the entire QNF plane into two distinct regions, 
where the QNFs of Kerr BHs for corotating and polar cases are located in the right region 
while those of Kerr BHs for counterrotating case and Taub-NUT BHs in the left one. 
We may refer to the two regions as two phases, i.e., the {\em Kerr-I phase} and {\em Taub-NUT {(or {Kerr-II})} phase}.
In other words, we think that Kerr and Taub-NUT BHs are two different states of Schwarzschild BHs in the complex plane of QNFs, 
and regard Fig.\ \ref{fig:zeeman-splitting} as a dynamical phase diagram in which the QNFs of Schwarzschild BHs represent a coexistence line. 
The QNFs of Kerr BHs  for corotating and polar cases are located in the right side of the coexistence line,
 while the QNFs of Kerr BHs for counterrotating case and Taub-NUT BHs in the left side. 
We may conclude that the above correlations of QNFs  (QNMs) are closely connected to the NJA among Schwarzschild, Kerr, and Taub-NUT BHs.
In other words, we may speculate that the connections to the NJA give rise to the correlation of QNFs, which may be referred to as the {\em Schwarzschild/Kerr/Taub-NUT (SKT) correspondence}, as shown in  
Fig.\ \ref{fig:zeeman-splitting}.

The light green curve depicts the variation of QNFs with respect to electric charge $Q$ for RN BHs. 
It is evident that the geometric characteristic of the QNFs of RN BHs is distinct from that of the other three BHs because RN BHs are not a member of the BHs connected by the NJA as depicted by Fig.~\ref{fig:relation}. 

The structure illustrated in Fig.~\ref{fig:zeeman-splitting}, consisting of the red, blue, and purple curves, is referred to as the Zeeman splitting of Kerr BHs~\cite{Onozawa:1996ux,Berti:2004md,Berti:2009kk}. 
This term is used because it bears some resemblance to the Zeeman effect observed in atomic physics. 
Apart from the splitting of Kerr BHs with different orbits, Fig.~\ref{fig:zeeman-splitting} provides additional information from the perspective of the NJA:
\begin{enumerate}
\item Schwarzschild BHs can be regarded as the seed of both Kerr and Taub-NUT BHs with respect to the NJA, 
and can be seen as the original state prior to the splitting. The 
NJA can be regarded as an ``external field'', similar to a magnetic field in atomic physics, 
while the Kerr and Taub-NUT BHs represent an even-number splitting of Schwarzschild BHs that is similar to an even-number energy level's splitting in atomic physics.
\item The QNFs of Taub-NUT BHs exhibit a well-defined splitting pattern, 
whereas the QNFs of RN BHs intersect with the QNFs of Kerr BHs' corotating orbit. 
This behavior may be connected to the NJA. 
Furthermore, the NJA produces distinct effects to different transformations: 
The transformatiom from Schwarzschild to Taub-NUT BHs results in a redshift\footnote{It means a decreasing of real parts, while the blueshift means an increasing of real parts.}
 of QNFs of Taub-NUT BHs compared to those of Schwarzschild BHs, 
whereas that from Schwarzschild to Kerr BHs leads to both redshifts (in the case of counterrotating orbits) 
and  blueshifts (in the case of corotating and polar orbits) when the even-number splitting of Schwarzschild BHs is caused by the NJA.
\item The Kerr BHs with different orbits and Taub-NUT BHs
can be categorized into two groups owing to the splitting of Schwarzschild BHs by the NJA. 
One group includes the Kerr BHs with corotating orbits and Taub-NUT BHs because they 
correspond to long-lived states where their Lyapunov exponents go to zero as $a\to M$ and $N\to \infty$, respectively. 
The other group consists of the Kerr BHs with counterrotating and polar orbits, 
where the Lyapunov exponents approach finite values as the parameter $a$ approaches $M$.
\item The Kerr and Taub-NUT BHs can be regarded as two special cases of Kerr-Taub-NUT BHs with the rotation parameter and NUT charge, 
while the RN and Kerr BHs can be seen as two special cases of  Kerr-Newman BHs with the electric charge and rotation parameter. Thus,
Fig.~\ref{fig:zeeman-splitting} also shows the QNFs of two families of BHs each of which has two parameters except mass. To be specific,
the QNF spectrum of Kerr-Taub-NUT BHs emerges well-splitting, where 
the Kerr and Schwarzschild BHs, as well as the Taub-NUT and Schwarzschild BHs, are related to each other through the NJA. 
On the other hand, for Kerr-Newman BHs, the QNF spectrum does not present the same splitting as that in Kerr-Taub-NUT BHs, but the crossing of spectral curves of RN and Kerr BHs (corotating case), where the RN and Schwarzschild black holes are not related through the NJA. This indicates that the QNF spectrum of a two-parameter family of BHs related to the NJA is different from that of a two-parameter family of BHs unrelated to the NJA.
However, it is important to note that RN BHs are classified under the Kerr-Newman-Taub-NUT family~\cite{Wu:2023eml,Ghezelbash:2021lcf,Yang:2023hll} by the NJA. In this context, RN BHs serve as the seed, from which the RN-Taub-NUT and Kerr-Newman BHs are correlated through the NJA. Consequently, the Kerr-Newman-Taub-NUT class exhibits a clear splitting, as illustrated in Fig.~\ref{fig:kntn}.

\begin{figure}[!ht]
     \centering
        \includegraphics[width=.6\textwidth]{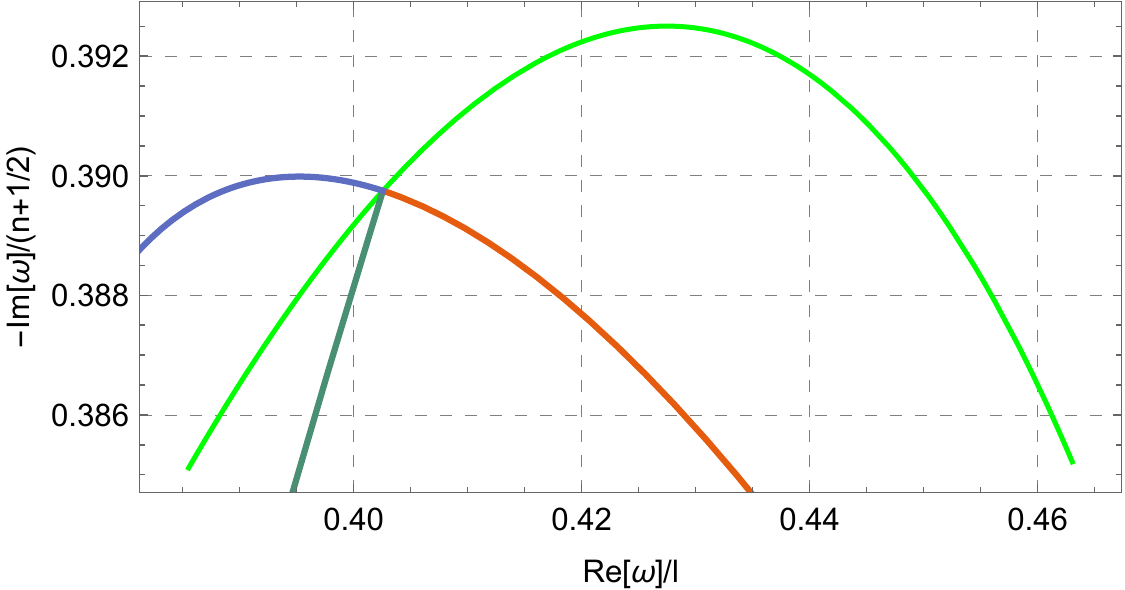}
      \captionsetup{width=.9\textwidth}
       \caption{QNMs of the Kerr-Newman-Taub-NUT class. The light green curve indicates the QNFs of RN BHs, the dark green curve represents the QNFs of RN-Taub-NUT BHs, and the red and blue curves correspond to the QNFs of the corotating and counterrotating Kerr-Newman BHs, respectively.}
        \label{fig:kntn}
\end{figure}
\end{enumerate}

\begin{figure}[!ht]
     \centering
     \begin{subfigure}[b]{0.45\textwidth}
         \centering
         \includegraphics[width=\textwidth]{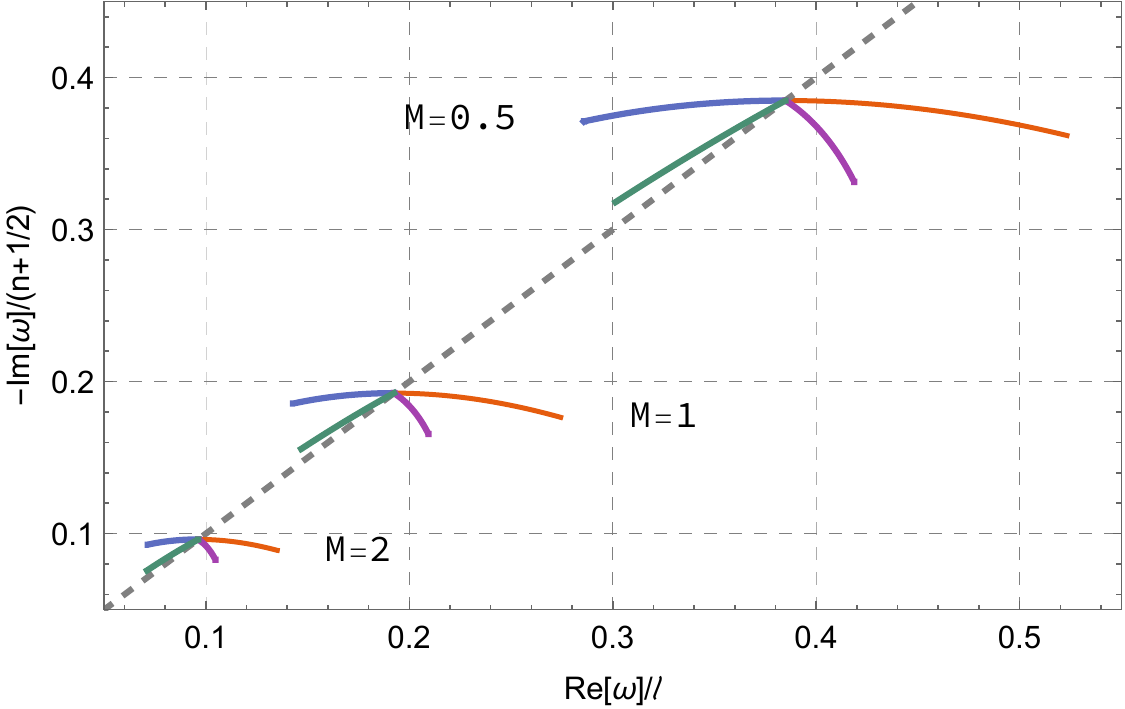}
         \caption{Schwarzschild BHs as the seed of the NJA.}
         \label{fig:zeeman-splitting2}
     \end{subfigure}
     \begin{subfigure}[b]{0.45\textwidth}
         \centering
         \includegraphics[width=\textwidth]{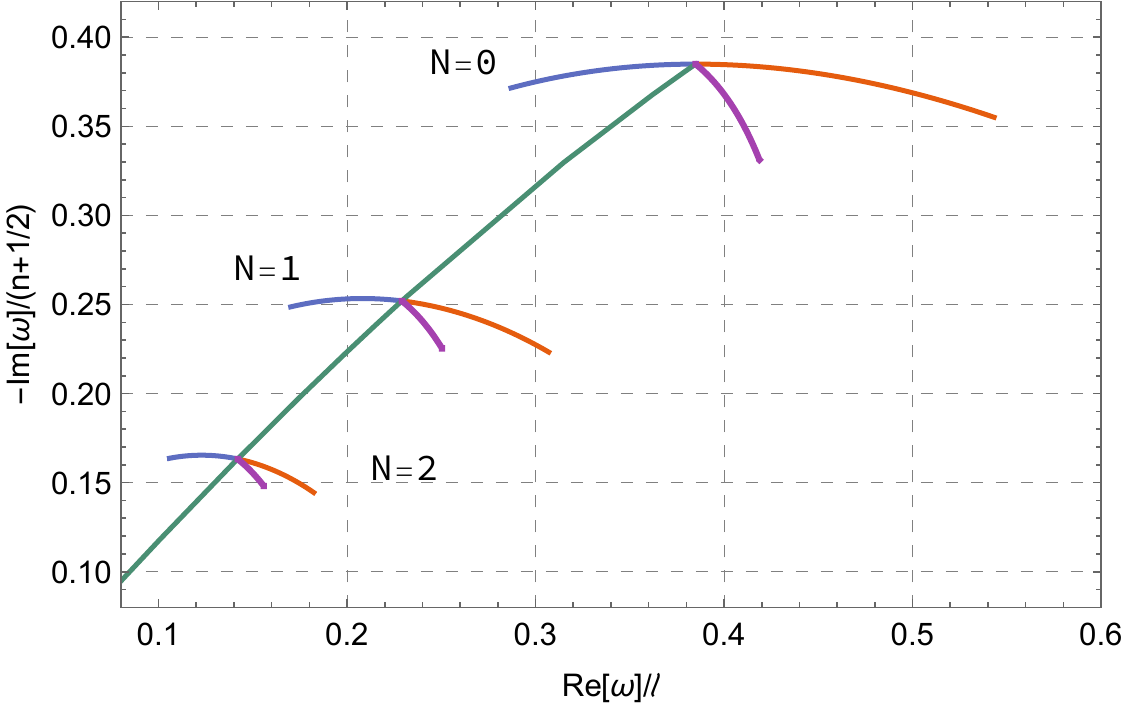}
         \caption{Taub-NUT BHs as the seed of the NJA.}
         \label{fig:zeeman-splitting3}
     \end{subfigure}
      \captionsetup{width=.9\textwidth}
       \caption{The Zeeman splitting of QNFs for two different seeds of the NJA.}
        \label{fig:zeeman-splitting23}
\end{figure}

Figure \ref{fig:zeeman-splitting23} depicts the Zeeman splitting of QNFs with different seeds of the NJA. Fig.~\ref{fig:zeeman-splitting2}, an extension of Fig.~\ref{fig:zeeman-splitting} with $M=0.5, 1, 2$, illustrates the splitting of Kerr and Taub-NUT BHs, starting with Schwarzschild BHs as the seed. On the other hand, Fig.~\ref{fig:zeeman-splitting3} depicts the splitting of Kerr-Taub-NUT BHs with respect to Taub-NUT BHs as the seed of the NJA.

The splittings of QNFs shown in Fig.\ \ref{fig:zeeman-splitting2} [Fig.\ \ref{fig:zeeman-splitting3}]  
also exhibit a scaling relationship with respect to the parameter $M$ ($N$). 
Specifically,  Fig.~\ref{fig:zeeman-splitting2} suggests that we can normalize the QNFs, i.e., we can make the QNFs dimensionless using the parameter $2M$ as a normalization factor, 
while Fig.~\ref{fig:zeeman-splitting3} indicates that the parameter $N$ can also serve as a normalization factor.
In other words, setting $M$ as $1/2$ in Fig.~\ref{fig:zeeman-splitting2} or $N$ as $1$ in Fig.~\ref{fig:zeeman-splitting3} yields the same results as the dimensionless treatment,
while the comparison with the cases of $M=1, 2$ in Fig.~\ref{fig:zeeman-splitting2} or $N=0, 2$ in Fig.~\ref{fig:zeeman-splitting3} gives the ``conformal" structure of the phase diagram.
Moreover, Fig.~\ref{fig:zeeman-splitting3}, in which the Lorentzian signature $(-+++)$ is adopted, exhibits more information than that in the Kleinian signature $(--++)$ or Euclidean signature $(++++)$.
Since
the self-dual Taub-NUT BHs can be transformed into the self-dual Kerr-Taub-NUT BHs by the coordinate transformations in the Kleinian and Euclidean signatures~\cite{Crawley:2021auj,Crawley:2023brz,Guevara:2023wlr},
one may expect that the spectra of Taub-NUT BHs and Kerr-Taub-NUT BHs have a one-to-one correspondence.

\subsection{General photon spheres}

We now turn our attention to the eikonal QNMs emanating from a general photon sphere, i.e., the photon sphere that is neither fully equatorial nor fully polar. 
Our objective is to investigate whether the QNM curves of a Kerr BH with a nonzero Carter’s constant will intersect with those of a Taub-NUT BH. The radius of a general photon sphere is given by~\cite{Teo:2003ltt,Gralla:2019ceu}
\begin{equation}
r^{\rm Ker}_{c}=M + 2 M \Delta_\zeta \cos\left[ \frac{1}{3} \cos^{-1}\left(\frac{1-a^2/M^2}{\Delta_\zeta^3}\right)\right],
\end{equation}
and the newly appeared parameters are defined by
\begin{equation}
\Delta_\zeta = \sqrt{1-\frac{a(a+\zeta)}{3M^2}},\qquad
\zeta= \frac{L_z}{E}.
\end{equation}
The unstable and non-equatorial null geodesics necessitate the Carter’s constant ${\cal Q}$ being greater than zero. This implies that the radius of general photon spheres is bounded by the corotating and  counterrotating radii of equatorial photon spheres, see Eq.~\eqref{eq:equatorial-ring},
\begin{equation}
r^{\rm Ker}_{+}<r^{\rm Ker}_{c}<r^{\rm Ker}_{-}.\label{rKBHnoneq}
\end{equation}
This condition further constrains $\zeta$ as follows:
\begin{equation}
\zeta_-<\zeta < \zeta_+,
\end{equation}
where $\zeta_\pm$ are defined as
\begin{equation}
\zeta_\pm=\frac{a^2 M+a^2 r^{\rm Ker}_{\pm}-3 M (r^{\rm Ker}_{\pm})^2+(r^{\rm Ker}_{\pm})^3}{a (M-r^{\rm Ker}_{\pm})}.
\end{equation}

On the other hand, ${\cal Q}>0$  indicates that the geodesics oscillate between two turning points $\theta_\pm$~\cite{Gralla:2019ceu},
\begin{equation}
\theta_\pm= \cos^{-1}\left(\mp\sqrt{u_+}\right),
\end{equation}
where $\theta_-\in(0,\pi/2)$ and  $\theta_+\in(\pi/2,\pi)$, and
\begin{equation}
u_\pm = \Delta_\theta+\sqrt{\Delta^2_\theta +\frac{\eta}{a^2}},\qquad
\Delta_\theta = \frac{1}{2} \left(
1-\frac{\eta+\zeta^2}{a^2}
\right).
\end{equation}
Here we use $\eta={\cal Q}/E^2$ to represent the reduced Carter's constant.

The eikonal QNMs can be calculated in terms of Eqs.~(1.1)-(1.2) in Ref.~\cite{Yang:2012he}, 
where the orbital and Lense-Thiring-precession frequencies are given by
\begin{equation}
\Omega_\theta(r^{\rm Ker}_{c}) = 2\pp /T_\theta(r^{\rm Ker}_{c}),\qquad
\Omega_{\rm prec}(r^{\rm Ker}_{c}) = \Omega_\theta(r^{\rm Ker}_{c}) \frac{\delta\varphi(r^{\rm Ker}_{c})}{2\pp} -({\rm sgn} L_z)\Omega_\theta(r^{\rm Ker}_{c}),
\end{equation}
and $T_\theta(r^{\rm Ker}_{c})$ and $\delta\varphi(r^{\rm Ker}_{c})$ can be represented by the elliptic functions~\cite{Teo:2003ltt,Gralla:2019ceu},
\begin{equation}
T_\theta(r^{\rm Ker}_{c}) = - \frac{4a^2 u_+}{\sqrt{- a^2 u_-}}E'\left(\frac{u_+}{u_-}\right)
+\frac{r \left[a^2 (2 M+r)-2 a \zeta  M+r^3\right]}{\Delta  }
\frac{2}{\sqrt{- a^2 u_-}}K\left(\frac{u_+}{u_-}\right),
\end{equation}
\begin{equation}
\delta \varphi(r^{\rm Ker}_{c}) = \frac{a (2 M r-a \zeta )}{\Delta } \frac{2}{\sqrt{- a^2 u_-}}K\left(\frac{u_+}{u_-}\right)
+\zeta\frac{2}{\sqrt{- a^2 u_-}}\Pi\left(u_+, \frac{u_+}{u_-}\right).
\end{equation}
Note that $r^{\rm Ker}_{c}$ takes the values in the range given by Eq.~(\ref{rKBHnoneq}).  Here $K\left(\frac{u_+}{u_-}\right)$, $E\left(\frac{u_+}{u_-}\right)$, and  $\Pi\left(u_+, \frac{u_+}{u_-}\right)$ stand for the first, second, and third classes of elliptic functions, and $E'\left(\frac{u_+}{u_-}\right)$ denotes the first-order derivative of the second class of elliptic functions with respect to ${u_+}/{u_-}$. The Lyapunov exponent takes the form,
\begin{equation}
\gamma_L =\left. \frac{\sqrt{2{\cal R}''(r)}\, \Delta}{\left[\frac{\partial {\cal R}}{\partial E}+\frac{\partial {\cal R}}{\partial {\cal Q}}\left(\frac{\dif {\cal Q}}{\dif E}\right)_{\rm BS}\right]}\right|_{r^{\rm Ker}_{c}},
\end{equation}
where ${\cal R}(r)$ is the radial potential, 
\begin{equation}
{\cal R}(r)=\left[E(r^2+a^2) - L_z a\right]^2
-\Delta\left[(L_z - a E)^2 + {\cal Q}\right],
\end{equation}
${\cal R}''(r)$ stands for the second-order derivative of ${\cal R}(r)$ with respect to $r$, and the subscript ``BS''  implies that the derivative, ${\dif {\cal Q}}/{\dif E}$, is determined by the angular Bohr-Sommerfeld condition~\cite{Yang:2012he},
\begin{equation}
\oint\dif\theta\sqrt{\Theta} = 2\pp \left(L-\abs{m}\right)
\end{equation}
where $\Theta={\cal Q}-\cos^2\theta(L_z^2/\sin^2\theta - a^2 E^2)$ is the angular potential and $L$ the angular momentum of test particles.

\begin{figure}[!ht]
     \centering
        \includegraphics[width=.6\textwidth]{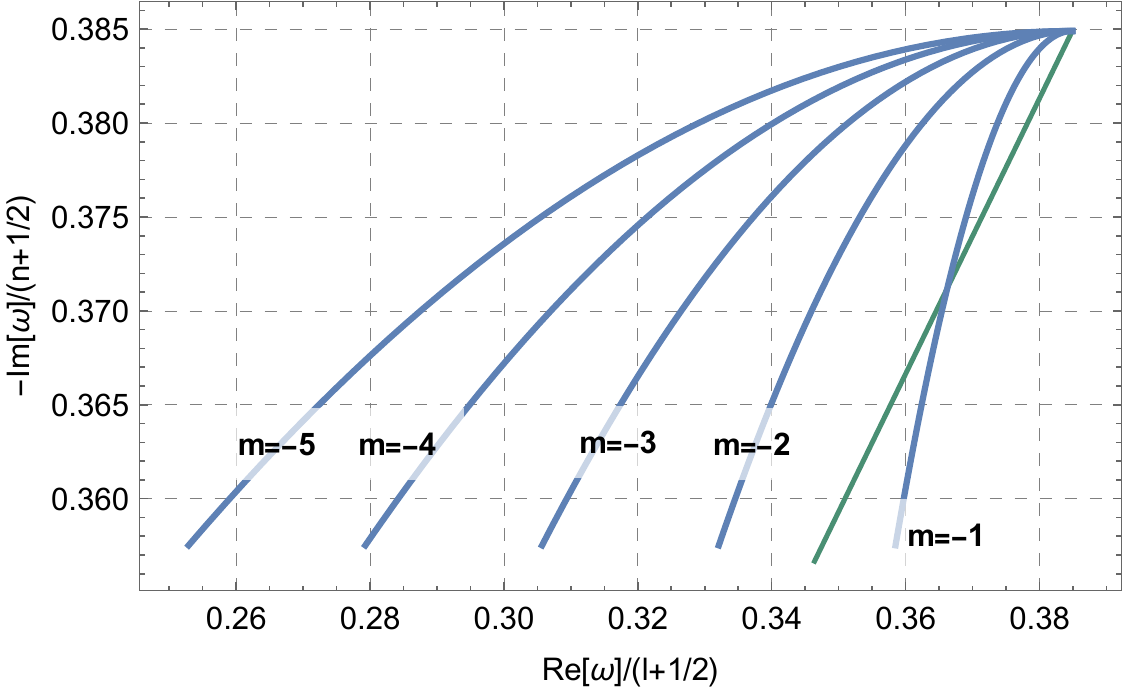}
      \captionsetup{width=.9\textwidth}
       \caption{QNMs correspond to Kerr's general photon spheres, where $l=5$ and $m=-1,...,-5$ are set.}
        \label{fig:generalLR}
\end{figure}

For a slow rotation, we plot the QNMs with respect to a negative azimuthal number\footnote{For a positive azimuthal number, the QNM curves of Kerr BHs go to the right-down direction, leading to no intersections with the QNM curves of Taub-NUT BHs.} $m$ for the Kerr's nonequatorial null geodesics (in blue)  
alongside the QNMs of  Taub-NUT BHs (in dark green), as shown in Fig.~\ref{fig:generalLR}. We observe that the Kerr QNM curve with $m=-1$ intersects with the Taub-NUT QNM curve.
In other words, a clear splitting occurs for $|m|/(l+1/2)\sim 1$, particularly for $m=\pm l$ and $l \gg 1$.

We end Sec.~\ref{sec:light-ring/qnms} by discussing the possibility of a phase transition that occurs in the SKT phases. 
When we plot the QNFs of Schwarzschild, Kerr, and Taub-NUT BHs in one diagram, 
it is natural to ask whether a transition between any two of the three phases occurs or not. 
The answer is negative, owing to the singularity or topological nature of the three BHs. In Sec.\ \ref{sec:curvature}, we have categorized the singularities of Kerr-Taub-NUT BHs in terms of the parameter space $(a, N)$. 
The curvature singularities of Schwarzschild, Kerr, and Taub-NUT BHs differ significantly, and the change of curvature singularities from one type to another implies~\cite{Borde:1996df} a change in topology. 
In other words, the difference in topology acts as a safeguard against phase transitions, preventing the occurrence of phase transitions.

Our next task is to acquire more accurate QNFs than the analytical ones from the light ring/QNMs correspondence. To this end, we need to perform numerical calculations which will be proceeded in the following three sections.

\section{Test-field perturbations and separations of variables}
\label{sec:separation}

In this section, we analyze the perturbations of two test fields with and without mass, i.e., a scalar field and a spinor field. Additionally, we demonstrate the process of variable separation in the Kerr-Taub-NUT spacetime.

\subsection{Scalar field perturbation}

The dynamics of a scalar field $\Psi$ that has a nonvanishing mass $m_{0}$ and is minimally coupled with gravity is governed by the Klein-Gordon equation in the Kerr-Taub-NUT spacetime,
\begin{equation}
\label{eq:klein-gordon}
    \left( \nabla^{\mu} \nabla_{\mu} -m_{0}^2 \right) \Psi=0,
\end{equation}
where the Greek superscripts and subscripts mean the temporal and spatial indices in the four-dimensional spacetime, and $\nabla_{\mu}$ stands for covariant derivative.
To perform the separation of variables in the above equation, we decompose~\cite{Teukolsky:1972my,Teukolsky:1973ha} the scalar field as 
\begin{equation}
    \Psi(t,r,\theta,\phi)=\me^{-\mi\omega t+\mi m \phi}R_{lm}(r)S_{lm}(\theta),
\end{equation} 
where  $\omega$ denotes the frequency of modes, $R_{lm}(r)$ the radial function,  $S_{lm}(\theta)$ the spheroidal angular function, $l$ the multipole  number, and $m$ the azimuthal number. By substituting this decomposition into the Klein-Gordon equation Eq.\ \eqref{eq:klein-gordon}, we obtain~\cite{Yang:2020iat,Yang:2023hll} the radial equation,
\begin{equation}
	\label{eq:scalar-radial}
	\Delta \frac{d}{d r}\left( \Delta \frac{d R_{lm}}{d r}\right)+\Big[ G^2+ (2  ma \omega-m_{0}^2 r^2-\lambda_{lm})\Delta \Big]R_{lm}=0,
\end{equation}
 and the angular equation, 
\begin{equation}
	\label{eq:scalar-angular}
	\frac{1}{\sin{\theta}}\frac{\dif}{\dif\theta}\left(\sin{\theta} \frac{\dif S_{lm}}{\dif \theta}\right)-\left[\frac{(2 N \omega \cos{\theta}-a \omega \sin^2{\theta}+m)^2}{\sin^2{\theta}}+m_{0}^2 \left(N+a \cos{\theta}\right)^2-\lambda_{lm}\right]S_{lm}=0,
\end{equation}
where $\lambda_{lm}$ serves as the separation constant,  $\Delta$ is given by Eq.~(\ref{defdelta}), and $G$ is defined by
\begin{equation}
G=\omega (r^2+a^2+N^2)-ma.
\end{equation}

\subsection{Spinor field perturbation}
\label{sec:dirac-TN}

In order to achieve the separation of spinor fields, we utilize~\cite{Chandrasekhar:1998kt} the Newman-Penrose formalism and express the Dirac equation in the following manner: 
\begin{equation}
\label{eq:dirac-np}
    \begin{split}
        (D+\epsilon-\rho)F_1 +(\bar{\delta}+\pi-\alpha) F_2 =\frac{1}{\sqrt{2}}\mi m_{e} G_1 ,\\
        (\triangle+\mu-\gamma)F_2+(\delta+\beta-\tau)F_1=\frac{1}{\sqrt{2}}\mi m_{e} G_2 ,\\
        (D+\bar{\epsilon}-\bar{\rho})G_2-(\delta +\bar{\pi}-\bar{\alpha})G_1=\frac{1}{\sqrt{2}}\mi m_{e} F_2 ,\\
        (\triangle+\bar{\mu}-\bar{\gamma})G_1-(\bar{\delta}+\bar{\beta}-\bar{\tau})G_2=\frac{1}{\sqrt{2}}\mi m_{e} F_1,
    \end{split}
\end{equation}
where the four-component spinor is written as $(F_1, F_2, G_1, G_2)$, $m_{e}$ is the mass of spinor fields, and a bar means the complex conjugate. Moreover, the three independent differential operators can be represented\footnote{Note the difference between $\triangle$ and $\Delta$, where the former is a triangle while the latter a capital Greek letter.} with the help of a null tetrad as follows:
\begin{equation}
D\coloneqq l^\mu \partial_\mu,\qquad \triangle\coloneqq n^\mu \partial_\mu,\qquad \delta \coloneqq m^\mu \partial_\mu,
\end{equation} 
and these Greek letters, $(\alpha, \beta, \gamma, \epsilon, \mu, \pi, \rho, \tau)$, denote the coefficients of spinor components. For the details of spinor fields in a curved spacetime, see Ref.~\cite{Collas:2018jfx}.

If the following null tetrad is applied,\footnote{This tetrad reduces to the one for Schwarzschild BHs~\cite{Jing:2005dt} or Kerr BHs~\cite{Chandrasekhar:1976ap} when the corresponding parameter, $a$ or $N$, goes to zero.}
\begin{equation}
    \begin{split}
       & l^\mu=\frac{1}{\Delta}\left\{r^2+a^2+N^2,\Delta,0,a\right\},\\ 
       & n^\mu = \frac{1}{2 \Sigma}\left\{r^2+a^2+N^2,-\Delta,0,a\right\},\\
       & m^\mu = \frac{1}{\sqrt{2} \Sigma_1}\left\{\mi (a \sin{\theta}-2 N \cot{\theta}),0,1,\mi \csc{\theta} \right\}, 
\end{split}
\label{eq:tetrad}
\end{equation}
we compute the nonvanishing coefficients of spinor components, 
\begin{equation}
    \begin{split}
    & \alpha=\pi-\bar{\beta},
    \qquad
    \beta=\frac{\cot{\theta}}{2 \sqrt{2} \Sigma_1},
    \qquad  
    \gamma=\mu+\frac{1}{4}\frac{\text{d} \Delta}{\text{d}r}, 
    \qquad
    \mu=-\frac{\Delta}{2 \Sigma \Sigma_2},
    \\
    &\pi=\frac{\mi a \sin{\theta}}{\sqrt{2}\Sigma_2^2},\qquad
     \rho= -\frac{1}{\Sigma_2},\qquad
    \tau=-\frac{\mi a \sin{\theta}}{\sqrt{2}\Sigma}.
    \end{split}
\label{eq:spin coeff}
\end{equation}
Further, substituting  
the following ansatz \cite{Chandrasekhar:1976ap} into Eq.\ \eqref{eq:dirac-np}, 
\begin{equation}
\begin{split}
    F_1=\frac{R_{-\frac{1}{2}}(r) S_{-\frac{1}{2}}(\theta)}{\Sigma_2} \me^{ -\mi\omega t +\mi m \phi},&\qquad
    F_2=R_{+\frac{1}{2}}(r) S_{+\frac{1}{2}}(\theta) \me^{ -\mi \omega t +\mi m \phi},\qquad\\
    G_1=R_{+\frac{1}{2}}(r) S_{-\frac{1}{2}}(\theta) \me^{ -\mi \omega t +\mi m \phi},&\qquad
    G_2=\frac{R_{-\frac{1}{2}}(r) S_{+\frac{1}{2}}(\theta)}{\Sigma_1} \me^{-\mi \omega t +\mi m \phi},
\end{split}
\end{equation}
we simplify the Dirac equation Eq.~(\ref{eq:dirac-np}) to be
\begin{subequations}
\begin{equation}
\label{eq:sep-dirac-radial}
    \begin{split}
       \mathcal{D}_0 R_{-\frac{1}{2}} & =(\lambda+\mi m_{e} r)R_{+\frac{1}{2}},\\
       \sqrt{\Delta} \mathcal{D}^{\dag}_{0} \left(\sqrt{\Delta}R_{+\frac{1}{2}}\right) & = \left(\lambda -\mi m_{e} r \right) R_{-\frac{1}{2}},
    \end{split}
\end{equation}   
\begin{equation}
\label{eq:sep-dirac-ang}
    \begin{split}
       \mathcal{L}_{\frac{1}{2}}S_{+\frac{1}{2}} & =\Big[-\lambda +m_{e}(a \cos{\theta}+N)  \Big]S_{-\frac{1}{2}},\\
       \mathcal{L}_{\frac{1}{2}}^{\dag}S_{-\frac{1}{2}} & =  \Big[\lambda + m_{e}(a\cos{\theta}+N)\Big]S_{+\frac{1}{2}},
    \end{split}
\end{equation}
\end{subequations}
where $\lambda$ is the separation constant and the differential operators take~\cite{Jing:2005dt} the forms,
\begin{equation}
\begin{split}
& \mathcal{D}_{k}=\frac{\partial }{\partial r}-\frac{\mi K}{\Delta}+\frac{k}{\Delta}\frac{\text{d$\Delta$}}{\text{d$r$}},\qquad \mathcal{D}^{\dag}_{k}=\frac{\partial }{\partial r}+\frac{\mi K }{\Delta}+\frac{k}{\Delta}\frac{\text{d$\Delta$}}{\text{d$r$}},\\
& \mathcal{L}_{k}=\frac{\partial}{\partial \theta } -Q + k \cot\theta,\qquad
\mathcal{L}^{\dag}_{k}=\frac{\partial}{\partial \theta } + Q + k \cot\theta,
\end{split}
\end{equation}
where $k$ denotes the spin of fields, for instance, $k=0$ for scalar fields, $k=1/2$ for spinor fields, etc, and the factors $K$ and $Q$ are defined by
\begin{equation}
K=\omega(r^2+a^2+N^2)-ma,\qquad
Q=\omega(a \sin{\theta}-2 N \cot{\theta})-\frac{m}{\sin{\theta}}.
\end{equation}  
We note that Eq.~\eqref{eq:sep-dirac-radial} describes
the radial equations and Eq.~\eqref{eq:sep-dirac-ang} the angular ones, where such a formulation is known~\cite{Chandrasekhar:1976ap,Page:1976jj} as the Chandrasekhar–Page-like equations.

At last, by decoupling $R_{+\frac{1}{2}}(r)$ from $R_{-\frac{1}{2}}(r)$ in Eq.~\eqref{eq:sep-dirac-radial} and $S_{+\frac{1}{2}}(\theta)$ from $S_{-\frac{1}{2}}(\theta)$ in Eq.~\eqref{eq:sep-dirac-ang}, respectively, we obtain the completely separated radial and
angular equations,
\begin{equation}
	\label{eq:dirac-radial}
	\begin{split}
		& \frac{\dif }{\dif r}\left(\Delta \frac{\dif P_s}{\dif r}\right)+\left(\frac{2\mi s m_e \Delta }{\lambda-2\mi s m_{e} r}-\frac{1}{2}\frac{d\Delta}{ dr}\right)\frac{d P_s}{dr}+\\
		&+\left[\frac{K^2-2 \mi s(r-M) K }{\Delta} +2 \mi s \frac{\dif  K}{\dif r}
		-\frac{m_{e} K}{\lambda-2 \mi s m_{e} r}-\lambda^2-m_{e}^2 r^2\right] P_s=0,
	\end{split}
\end{equation}
and 
\begin{equation}
	\label{eq:dirac-angular}
	\begin{split}
		& \frac{1}{\sin{\theta}}\frac{\dif}{\dif \theta}\left( \sin{\theta} \frac{\dif S_s}{\dif \theta}\right)+\frac{a m_{e} \sin{\theta}}{m_{e}(a \cos{\theta}+N)-2 s}\frac{\dif S_s}{\dif\theta}+ \Bigg[m_{e}^2\left(a \cos{\theta} + N \right)-\lambda^2 \\
		&+\frac{1}{2}\left(\cot{\theta}-2 Q\right)\frac{am_{e}  \sin{\theta}}{m_{e}\left(a \cos{\theta}+N\right)-2 s}+\frac{1}{4}\left(\cot^2{\theta}-\frac{2}{\sin^2{\theta}}-4 Q^2-8 s\frac{\dif Q}{\dif \theta}\right) \Bigg]S_s=0,
	\end{split}
\end{equation}
where $s$ denotes the spin of spinor fields, 
$s=\pm \frac{1}{2}$, 
and the  radial functions are rewritten to be
\begin{equation}
P_{+\frac{1}{2}}(r)=\sqrt{\Delta(r)}\,R_{+\frac{1}{2}}(r),\qquad
P_{-\frac{1}{2}}(r)=R_{-\frac{1}{2}}(r).\label{radequP+-}
\end{equation}

\section{Eigenvalues of angular equations}
\label{sec:angular}

In this section, we address the eigenvalues associated with the Chandrasekhar-Page-like equations, i.e., the separated angular equations described by Eq.~\eqref{eq:dirac-angular}. 
However, as the solutions for Kerr BHs have already been discussed~\cite{Seidel:1988ue,Berti:2005gp}, we focus primarily on the case of Taub-NUT BHs by setting $a = 0$ in Eq.~\eqref{eq:dirac-angular}. 
Additionally, for the sake of simplicity in notations, we omit the subscript $1/2$ in the spheroidal angular functions, and just use $S_\pm$ instead.
Thus, we simplify Eq.~\eqref{eq:dirac-angular} to be
\begin{equation}
\label{eq:ang-wave-orig}
    \frac{1}{\sin\theta}
    \frac{\dif }{\dif \theta}
    \left(\sin\theta \frac{\dif S_\pm }{\dif \theta}\right)
    +\left\{\lambda ^2-\frac{1}{2}-\left[\frac{m}{\sin \theta}+\left(2 N \omega \mp\frac{1}{2}\right)\cot \theta \right]^2- N \left(m_e^2 N\mp 2\omega \right)\right\}
    S_\pm=0,
\end{equation}
where the angular parameter $\theta$ is bounded by $\theta\in[0,\pp)$.
Alternatively, we recast Eq.~\eqref{eq:ang-wave-orig} by the replacement
$x=\sin(\theta/2)$ with
$x\in[0,1]$, and then obtain
\begin{multline}
\label{eq:master}
    x \left(1-x^2\right) \frac{\dif }{\dif x}\left[x \left(1-x^2\right) \frac{\dif}{\dif x} S_\pm\right]
    -\frac{1}{4} S_\pm\Big[4 m^2-4 m \left(2 x^2-1\right) (4 N \omega \mp 1)\\
    +4 x^2 \left(x^2-1\right) \left(4 \lambda ^2-4 m_e^2 N^2-1\right)
    +16 N^2 \left(1-2 x^2\right)^2 \omega ^2\mp 8 N \omega +1\Big]=0.
\end{multline}
The two second-order differential equations depicted by Eq.~(\ref{eq:master}) have two regular singular points located
at $x=0$ and $x=1$, respectively,
indicating that the naive solutions without any boundary conditions will have the same singularities at $x=0$ and $x=1$.

Furthermore, we note the symmetry of spheroidal angular functions between $S_+$ and $S_-$, i.e., $S_+$ can be converted to $S_-$ by the following transformation,
\begin{equation}
\label{eq:symmetry}
    N\to -N,\qquad
    m\to -m.
\end{equation}
This symmetry can help us to simplify the process of solving Eq.~(\ref{eq:master}).

We solve Eq.~\eqref{eq:master} and give the solutions via the hypergeometric functions,
\begin{eqnarray}
\label{eq:sol-aswave}
    S_\pm &=&
\mathcal{C}_1 x^{-\frac{1}{2}\pm m\pm 2 N \omega } \left(1-x^2\right)^{-\frac{1}{4}\mp \frac{m}{2}\pm N \omega } \, _2F_1\left(a_1^\pm,b_1^\pm;c_1^\pm ;x^2\right)
   \nonumber \\ 
   & &+\; \mathcal{C}_2 x^{\frac{1}{2}\mp m\mp 2 N \omega } 
    \left(1-x^2\right)^{-\frac{1}{4}\mp \frac{m}{2}\pm N \omega } \, 
    _2F_1\left(a_2^\pm,b_2^\pm;c_2^\pm;x^2\right), 
\end{eqnarray}  
where $\mathcal{C}_{1}$ and $\mathcal{C}_{2}$ are two arbitrary constants and the parameters of hypergeometric functions in two branches of solutions take the following forms,
\begin{subequations}
    \begin{equation}
        a_1^\pm=\pm 2 N \omega -\sqrt{\lambda ^2-N^2 \left(m_e^2-4 \omega ^2\right)},\label{firsola}
    \end{equation}
    \begin{equation}
        b_1^\pm=\pm 2 N \omega  +\sqrt{\lambda ^2-N^2 \left(m_e^2-4 \omega ^2\right)},\label{firsolb}
    \end{equation}
    \begin{equation}
        c_1^\pm =
       \frac{1}{2}\pm m \pm 2 N \omega;
    \end{equation}
\end{subequations}
and
\begin{subequations}
    \begin{equation}
        a_2^\pm=\frac{1}{2}\mp m- \sqrt{\lambda ^2-N^2 \left(m_e^2-4 \omega ^2\right)},\label{secsola}
    \end{equation}
    \begin{equation}
        b_2^\pm=\frac{1}{2}\mp m+\sqrt{\lambda ^2-N^2 \left(m_e^2-4 \omega ^2\right)},\label{secsolb}
    \end{equation}
    \begin{equation}
        c_2^\pm =
        \frac{3}{2}\mp m-2 N \omega. 
    \end{equation}
\end{subequations}
The naive solutions Eq.~\eqref{eq:sol-aswave} have two possible singularities at $x=0$ and $x=1$.
Therefore, we have to restrict the parameter $\lambda$ in order to construct normalizable eigenstates.
This procedure provides us with eigenvalues of the spheroidal angular functions.

To this end, we at first consider the asymptotic behavior of $S_+$ in the limit of 
$x\to 0$,
\begin{equation}
    S_+\sim \mathcal{C}_1 x^{m+2 N \omega -\frac{1}{2}}
    +\mathcal{C}_2 x^{-m-2 N \omega +\frac{1}{2}},
\end{equation}
where we have used~\cite{NIST:DLMF} the asymptotic formulas of hypergeometric functions around $x=0$.
\begin{itemize}
\item If $m> 1/2-2 N \Re(\omega)$, the first branch of solutions is finite, whereas the second one is divergent, indicating that we have to eliminate the second one by setting $\mathcal{C}_2=0$ in Eq.~\eqref{eq:sol-aswave}.
\item If $m< 1/2-2 N \Re(\omega)$, we have to remove the first branch by setting $\mathcal{C}_1=0$ but retain the second one in Eq.~\eqref{eq:sol-aswave}.
\end{itemize}

Next, we turn to the study of asymptotic behaviors in the limit of $x\to 1$. Considering the asymptotic formulas of hypergeometric functions around $x=1$~\cite{NIST:DLMF},
we obtain
\begin{equation}
    S_+\sim 
    \mathcal{C}_1\frac{ (1-x)^{-\frac{m}{2}+N \omega -\frac{1}{4}}}{\Gamma (1-a_2^+) \Gamma (1-b_2^+)}   +
    \mathcal{C}_2\frac{ (1-x)^{-\frac{m}{2}+N \omega -\frac{1}{4}}}{\Gamma (1-a_1^+) \Gamma (1-b_1^+)}.\label{asymxto1S+}
\end{equation}
\begin{itemize}
\item For the first case, i.e., $m> 1/2-2 N \Re(\omega)$, which gives rise to $\mathcal{C}_2=0$, it is possible that the power of $(1-x)$ in the first branch of solutions is negative because of $m\in \mathbb{Z}$, leading to the divergence of this branch. 
Thus, in order to overcome such a divergence, 
we demand
\begin{equation}
   \frac{1}{\Gamma (1-a_2^+) \Gamma (1-b_2^+)}=0,
\end{equation}
which implies that either $1-a_2^+=-l$ or $1-b_2^+=-l$ according to the property of the Gamma functions, where $l\in \mathbb{Z}^+$.
Using Eq.~(\ref{secsola}) or Eq.~(\ref{secsolb}), we obtain that these two conditions ($1-a_2^+=-l$ and $1-b_2^+=-l$) result in a unique $\lambda_+$,
\begin{equation}
    \lambda^2_+=N^2 \left(m_e^2-4 \omega ^2\right)+l \left(l+2 m+1\right)+m(m+1)+\frac{1}{4}.\label{1stclamd+}
\end{equation}
\item For the second case, i.e., $m< 1/2-2 N \Re(\omega)$, which gives rise to  $\mathcal{C}_1=0$, it is possible that the power of $(1-x)$ in the second branch of solutions is negative owing to $m\in \mathbb{Z}$, leading to the divergence of this branch. Thus, in order to overcome such a divergence, see Eq.~(\ref{asymxto1S+}),
we require
\begin{equation}
	\frac{1}{\Gamma (1-a_1^+) \Gamma (1-b_1^+)}=0,
\end{equation}
which implies $1-a_1^+=-l$ or $1-b_1^+=-l$, where $l\in \mathbb{Z}^+$.
Using Eq.~(\ref{firsola}) or Eq.~(\ref{firsolb}), we deduce a unique  $\lambda_+$,
\begin{equation}
    \lambda^2_+ =N^2 m_e^2+l(l-4 N \omega +2)-4 N \omega +1.\label{2ndclamd+}
\end{equation}
\end{itemize}

Moreover, considering the symmetry given by  Eq.~\eqref{eq:symmetry}, we establish the relationship between $\lambda_+$ and $\lambda_-$ as follows:
\begin{equation}
  \lambda^2_+  \xrightarrow{\substack{N\to -N\\m\to -m}} \lambda^2_-.
\end{equation}
Consequently, $\lambda_-$ takes the forms in the following two cases,
\begin{itemize}
\item If $m<-1/2-2 N \Re(\omega )$, we have
\begin{equation}
	\lambda^2_-=
	N^2 \left(m_e^2-4 \omega ^2\right)+l \left(l-2 m+1\right)+m (m-1) +\frac{1}{4},\label{1stclamd-}
\end{equation}
\item If $m>-1/2 -2 N \Re(\omega )$, we have
\begin{equation}
	\lambda^2_- =N^2 m_e^2+l \left(l+4 N \omega +2\right)+4 N \omega +1.\label{2ndclamd-}
\end{equation}
\end{itemize}

As the separation constant of angular and radial functions, $\lambda_\pm$ will be determined after we solve the radial equations and give the values of $\omega$, the QNFs of spinor field perturbations. 
It is possible that $\lambda_\pm$ are complex if $\omega$ is complex, 
see  Eqs.~(\ref{1stclamd+}), (\ref{2ndclamd+}), (\ref{1stclamd-}), and (\ref{2ndclamd-}).
In addition, we notice that $\lambda_+$ is irrelevant to $m$ when $m$ is small, i.e., $m< 1/2-2 N \Re(\omega)$, see Eq.~(\ref{2ndclamd+}), and that $\lambda_-$ is irrelevant to $m$ when
$m$ is large, i.e., $m>-1/2 -2 N \Re(\omega )$, see Eq.~(\ref{2ndclamd-})
.

\section{Relations in spectra of quasinormal frequencies}
\label{sec:relation}

In this section, we employ the continued fraction method to calculate the QNFs by solving the radial equations, Eq.~\eqref{eq:scalar-radial} and Eq.~\eqref{eq:dirac-radial}, which correspond to scalar and spinor field perturbations, respectively, in the Kerr-Taub-NUT spacetime. Subsequently, we analyze the relationships among the QNFs of Schwarzschild, Kerr, and Taub-NUT BHs.

\subsection{Continued fraction method}\label{conframeth}

The continued fraction method, also known as Leaver's method, is considered to be a more accurate approximation compared to others. It was initially introduced~\cite{Leaver:1985ax} by Leaver for massless field perturbations, and was later improved~\cite{Nollert} by Nollert. When we utilize the Leaver method to calculate QNFs for certain models, such as Schwarzschild and Kerr BHs, we usually encounter a three-term recurrence relation:
\begin{equation}
	\alpha_{n} a_{n-1}+\beta_{n} a_{n}+ \gamma_{n} a_{n+1} =0, \qquad n=1,2,\ldots,
	\label{eq:recurrence2}
\end{equation}
whose initial one is special and just contains two terms, 
\begin{equation}
    \alpha_{0} a_1 + \beta_{0} a_{0} =0,
\label{eq:recurrence1}
\end{equation}
where $a_n$'s are coefficients of series solutions, 
and $\alpha_n$'s, $\beta_n$'s, and $\gamma_n$'s are coefficients of the above recurrence relations.
The three-term recurrence relation gives the most fundamental scenario, but in certain models, we may encounter more-term recurrence relations, such as a four-term recurrence relation or even over four-term ones. When dealing with recurrence relations involving more than three terms, we can utilize the Gaussian elimination to simplify them and convert them back to a three-term recurrence relation. For more specific treatments, see Ref.~\cite{Leaver:1990zz}.

The three-term recurrence relation, as shown in Eq.\ \eqref{eq:recurrence2},
can be reformulated as a continued fraction, 
\begin{equation}
\label{eq:cond-infty}
    \frac{a_{n+1}}{a_{n}}=-\frac{\gamma_{n+1}}{\beta_{n+1}-\frac{\alpha_{n+1}\gamma_{n+2}}{\beta_{n+2}-\frac{\alpha_{n+2}\gamma_{n+3}}{\beta_{n+3}-\cdots}}}.
\end{equation}
For the case of $n=0$, we have
\begin{equation}
\label{eq:right-recurrence}
    \frac{a_{1}}{a_{0}}=-\frac{\gamma_{1}}{\beta_{1}-\frac{\alpha_{1}\gamma_{2}}{\beta_{2}-\frac{\alpha_{2}\gamma_{3}}{\beta_{3}-\cdots}}},
\end{equation}
and then replacing $a_1/a_0$ by Eq.\ \eqref{eq:recurrence1},
we derive an infinite continued fraction,
\begin{equation}
    0=\beta_{0}-\frac{\alpha_{0}\gamma_{1}}{\beta_{1}-\frac{\alpha_{1}\gamma_{2}}{\beta_{2}-\frac{\alpha_{2}\gamma_{3}}{\beta_{3}-\cdots}}}.
\label{eq:inf-fraction}
\end{equation}
Since the coefficients $\alpha_n$, $\beta_n$ and $\gamma_n$ are functions of $\omega$,
the most stable roots of Eq.~\eqref{eq:inf-fraction} represent frequencies, which are just the QNFs. 
In other words, the QNFs, representing the stability of black holes, correspond to minimum negative imaginary parts solved from Eq.~\eqref{eq:inf-fraction}, for the details, see Refs.~\cite{Leaver:1985ax,Nollert}. In the subsequent calculations, we use finite steps of continued fractions and perform 15 times of iterations (equivalent to $15$ steps in Eq.~\eqref{eq:inf-fraction}).

In Schwarzschild and Kerr BHs, the recurrence relations have been successfully derived for massless and massive scalar field perturbations, see Refs.~\cite{Leaver:1985ax, Konoplya:2004wg, Konoplya:2006br}, and they have also been computed for massless spinor field perturbation, see Refs.~\cite{Jing:2005dt, Jing:2005pk}. For massive spinor field perturbation,
the recurrence relations have been calculated~\cite{Konoplya:2017tvu} specifically for Kerr BHs. Therefore, we focus on the recurrence relations for (massless and massive) scalar and spinor field perturbations in Taub-NUT BHs.

In the case of Taub-NUT BHs, the boundary conditions for the radial function $R_{lm}(r)$ ($P_s(r)$) in  Eq.~\eqref{eq:scalar-radial} [Eq.~\eqref{eq:dirac-radial}] with the rotation parameter $a=0$ can be represented as 
\begin{eqnarray}
\label{eq:boundary}
R_{lm}(r)\; (\text{or} \; P_{s}(r))  \sim \left\{
\begin{array}{ll} \left(r-r_{+}\right)^{-\mi \omega r_{+} - \epsilon}, &
~~~~r\rightarrow r_+, \\
     \me^{\mi \chi  r}  r^{ \mi (\chi^2 + \omega^2)/(2 \chi)}, & ~~~~     r\rightarrow +\infty,
\end{array} \right.
\end{eqnarray}
where $\chi=\sqrt{\omega^2-m_{0}^2}$ for a scalar field perturbation and $\chi=\sqrt{\omega^2-m_{e}^2}$  for a spinor field perturbation, respectively. The parameter $\epsilon$ can take three values: $0$ and $\pm1/2$, representing a scalar field and a spinor field with the spin $1/2$ or $-1/2$, respectively.
Therefore, the solution of the radial equation for a scalar (spinor) field perturbation in Taub-NUT BHs can be expressed as follows:
\begin{equation}
\label{eq:ansatz}
R_{lm}(r)\;(\text{or}\; P_{s}(r))=\me^{\mi \chi r} \left(r- r_{-}\right)^{\mi (\chi^2+\omega^2)/(2 \chi)+\mi \omega r_{+} - \epsilon} \left(r-r_{+}\right)^{-\mi \omega r_{+} +\epsilon} \sum^{\infty}_{k=0} a_k \left(\frac{r-r_{+}}{r-r_{-}}\right)^k,
\end{equation}
where 
$r_-$ and $r_+$ stand for the inner and outer horizons of Taub-NUT BHs, respectively.  
By substituting  Eq.~\eqref{eq:ansatz} into Eq.~\eqref{eq:scalar-radial} [Eq.~\eqref{eq:dirac-radial}] for a scalar (spinor) perturbation in Taub-NUT BHs, we obtain a four-term recurrence relation of $a_k$,
\begin{equation}
    \alpha_{k}^{\epsilon} a_{k-1}+\beta_{k}^{\epsilon} a_{k}+ \gamma_{k}^{\epsilon} a_{k+1} +\delta_{k}^{\epsilon} a_{k+2}=0,
\label{eq:NUT-recurrence}
\end{equation}
The coefficients of recurrence relations, $\alpha_{k}^{\epsilon}$, $\beta_{k}^{\epsilon}$, $\gamma_{k}^{\epsilon}$, and  $\delta_{k}^{\epsilon}$, which can be determined analytically, are moved to Appendix \ref{app:rec-coef} owing to their tedious expressions.

\subsection{Numerical results}

As stated in Sec.~\ref{conframeth}, the calculations have primarily been made in the Taub-NUT spacetime. 
Specifically, we have considered the perturbations under a scalar and spinor fields with and without mass. 
Now we want to demonstrate the SKT correspondence in terms of QNFs, for the definition of such a correspondence, see Sec.~\ref{sec:num-qnf}.

\begin{figure}[!ht]
     \centering
     \begin{subfigure}{0.45\textwidth}
         \centering
     \includegraphics[width=\textwidth]{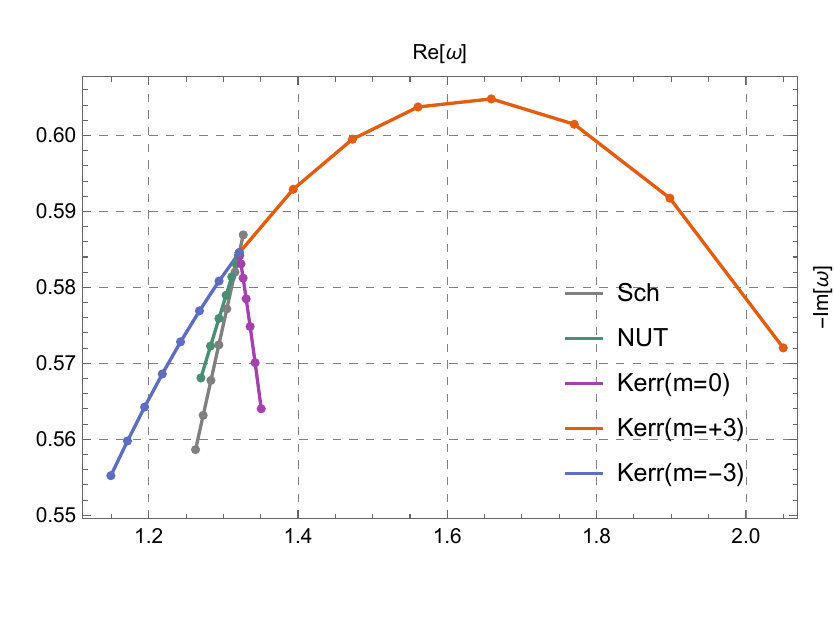}
         \caption{}
         \label{fig:scalar-massless}
     \end{subfigure}
     \begin{subfigure}{0.45\textwidth}
         \centering
         \includegraphics[width=\textwidth]{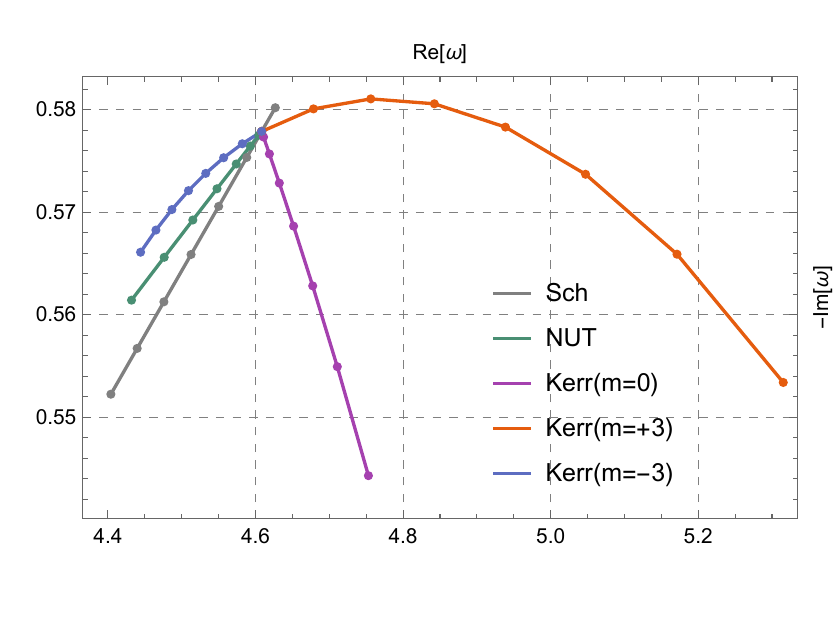}
         \caption{}
         \label{fig:dirac-massless}
     \end{subfigure}
      \captionsetup{width=.9\textwidth}
       \caption{Phase diagram of QNFs under a massless field perturbation, where $M=1/2$, $N=0, 0.038, 0.076, \cdots, 0.266$ for Taub-NUT BHs, $a=0, 0.02, 0.04, \cdots, 0.14$ for Kerr BHs with counterrotating orbits of $m=-3$, and $a=0, 0.05, 0.10, \cdots, 0.35$ for Kerr BHs with corotating orbits of $m=+3$ and polar orbits of $m=0$ are set. (a) Scalar field without mass. (b) Spinor field without mass. }
       \label{fig:massless}
\end{figure}

\begin{figure}[!ht]
	\centering
	\begin{subfigure}{0.45\textwidth}
		\centering
		\includegraphics[width=\textwidth]{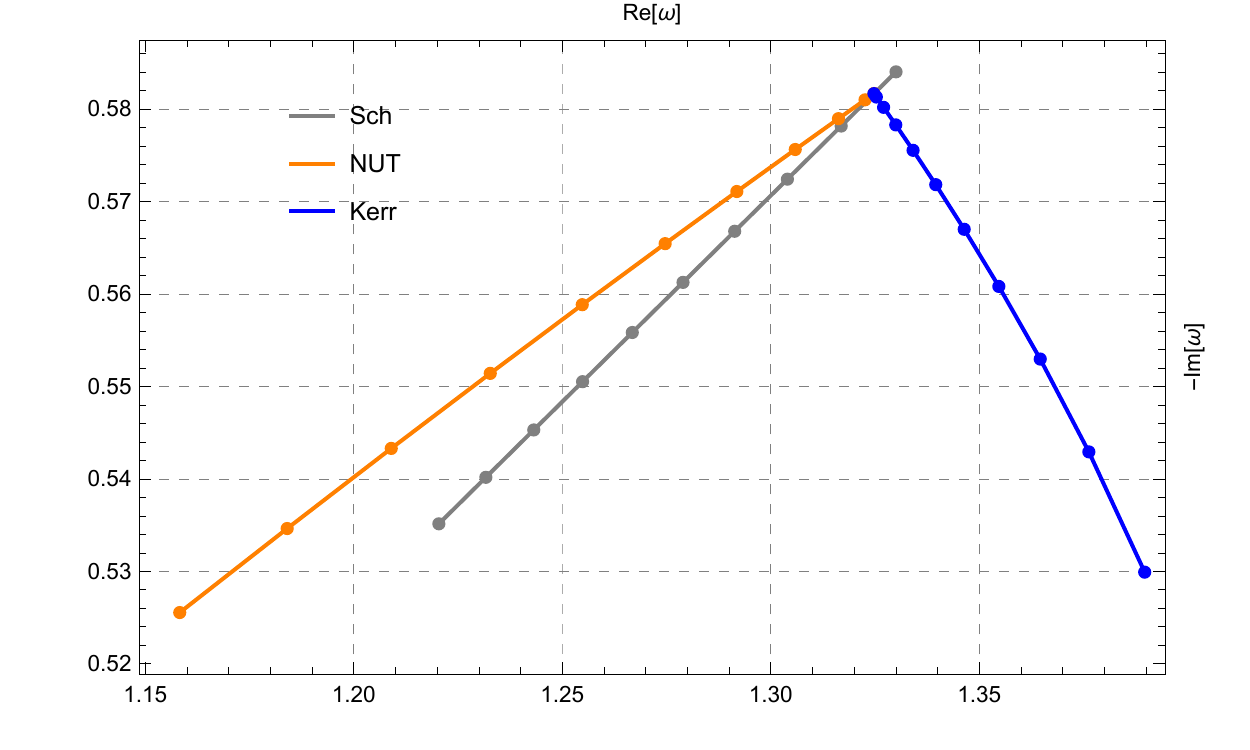}
		\caption{}
		\label{fig:scalar-massive}
	\end{subfigure}
	\begin{subfigure}{0.45\textwidth}
		\centering
		\includegraphics[width=\textwidth]{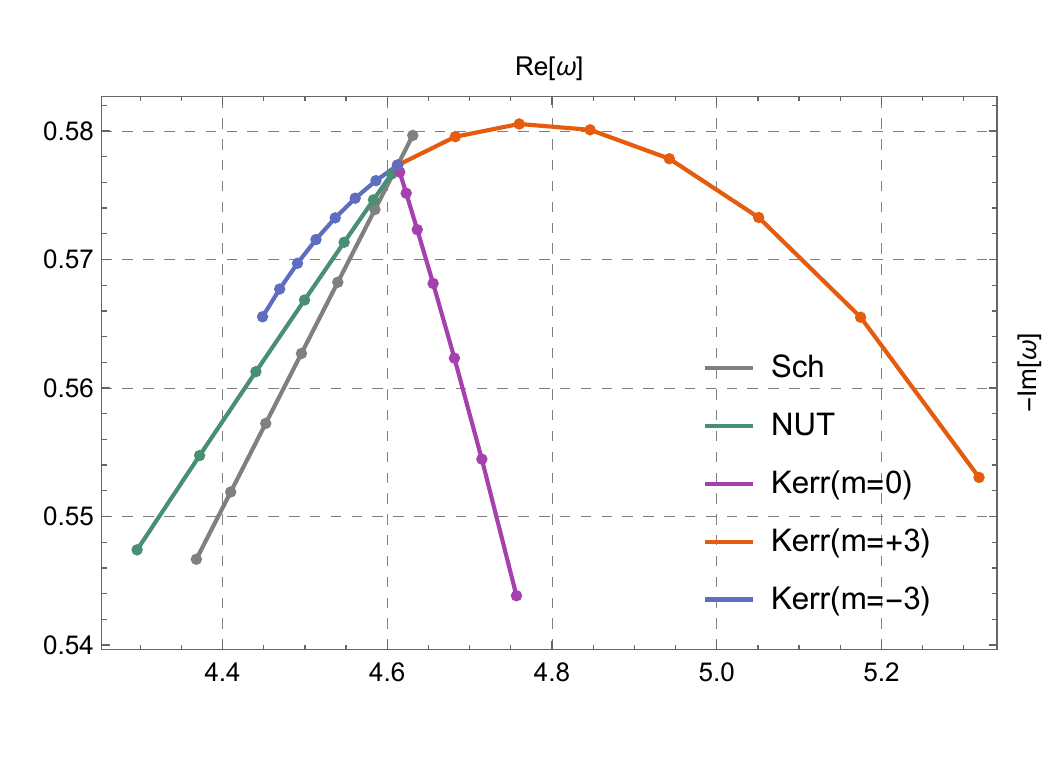}
		\caption{}
		\label{fig:dirac-massive}
	\end{subfigure}
	\captionsetup{width=.9\textwidth}
	\caption{Phase diagram of QNFs under a massive field perturbation, where $m_0=m_{e}=0.2$, $M=1/2$, $N=0, 0.038, 0.076, \cdots, 0.266$ for Taub-NUT BHs, $a=0, 0.02, 0.04, \cdots, 0.14$ for Kerr BHs with counterrotating orbits of $m=-3$, and $a=0, 0.05, 0.10, \cdots, 0.35$ for Kerr BHs with corotating orbits of $m=+3$ and polar orbits of $m=0$ are set. (a) Scalar field with mass. (b) Spinor field with mass.}
	\label{fig:massive}
\end{figure}

The primary findings of the SKT correspondence are illustrated in Fig.~\ref{fig:massless} and Fig.~\ref{fig:massive} for massless and massive field perturbations, respectively.
 
In Figs.~\ref{fig:massless} and \ref{fig:massive}, we assign $l = 3$ for the scalar case with and without mass, and we set $\lambda = 12$ for the spinor case with and without mass. 
The gray lines illustrate the changes in the QNFs of Schwarzschild BHs as the mass parameter $M$ varies from $0.498$ to $0.533$, where the step size is $0.005$ and then
eight equally spaced points are selected. 
The dark green curves represent the QNFs of Taub-NUT BHs, where we choose eight equally spaced points between $N = 0$ and $N = 0.266$ with a step size of $0.038$ and fix the mass at $M = 1/2$. 
The blue curves depict the changes in QNFs of Kerr BHs associated with counterrotating orbits with respect to the rotation parameter $a$, where we take eight equally spaced points from $a = 0$ to $a = 0.14$
with a step size of $0.02$ and fix the mass fixed $M = 1/2$ and the azimuthal number $m = -3$. 
The red (purple) curves represent the variations in the QNFs of Kerr BHs associated with corotating (polar) orbits with respect to 
the rotation parameter $a$, we take eight equidistant points from $a = 0$ to $a = 0.35$ with a step size of $0.05$ and fix the mass $M = 1/2$ and azimuthal number $m = 3$ ($m = 0$).
To sum up, Figs.~\ref{fig:scalar-massless} and \ref{fig:scalar-massive} display the results for a massless and massive scalar field perturbations, respectively, and Figs.~\ref{fig:dirac-massless} and \ref{fig:dirac-massive} show the results for a massless and massive spinor field perturbations, respectively. 
It is evident that both the massless case depicted in Fig.\ \ref{fig:massless} and the massive case shown in Fig.\ \ref{fig:massive} are in agreement with the earlier findings presented in Fig.\ \ref{fig:zeeman-splitting} obtained through the light ring/QNMs correspondence.

\begin{figure}[!ht]
	\centering
	\begin{subfigure}{0.45\textwidth}
		\centering
		\includegraphics[width=\textwidth]{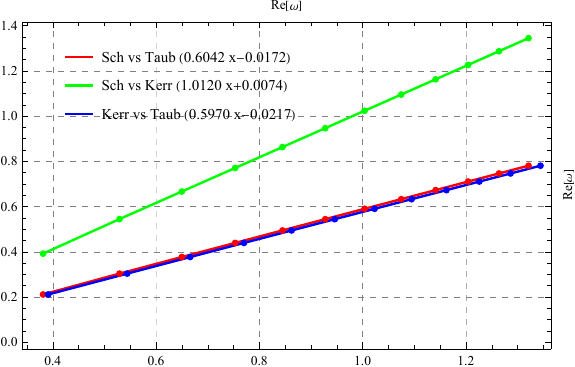}
		\caption{}
		\label{fig:scalar-massless-Re-Re-l-10}
	\end{subfigure}
	\begin{subfigure}{0.45\textwidth}
		\centering
		\includegraphics[width=\textwidth]{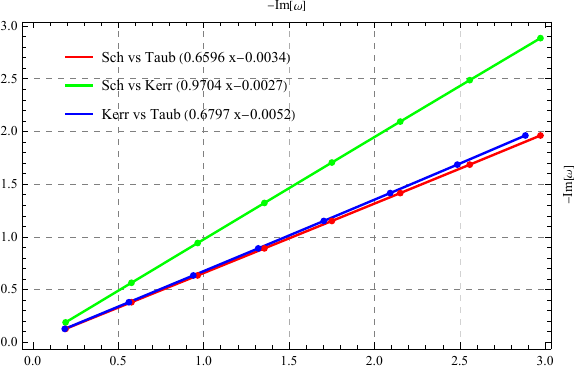}
		\caption{}
		\label{fig:scalar-massless-Im-Im-l-10}
	\end{subfigure}
	\captionsetup{width=.9\textwidth}
	\caption{QNFs of Schwarzschild, Kerr, and Taub-NUT  BHs under a massless scalar field perturbation. 
	The rotation parameter $a=0.25$ for Kerr BHs and the NUT charge $N=1$ for Taub-NUT BHs are set. Moreover, $m=0$ is set for Kerr BHs. (a) Real parts, $n=1$ and $\lambda=1, 2, \cdots, 12$. (b) Imaginary parts,  $l=10$ and $n=0, 1, \cdots, 7$.\label{fig:scalar-massless-slope}}
\end{figure}

\begin{figure}[!ht]
	\centering
	\begin{subfigure}{0.45\textwidth}
		\centering
		\includegraphics[width=\textwidth]{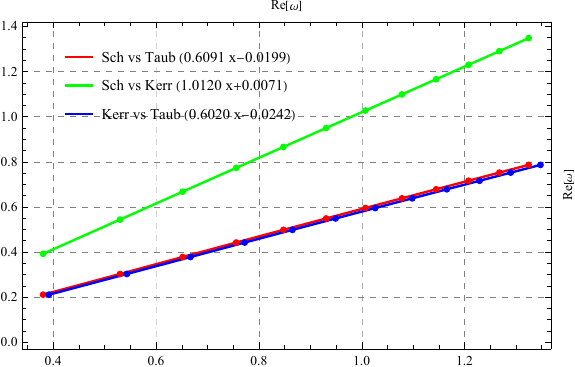}
		\caption{}
		\label{fig:scalar-massive-Re-Re-l-10}
	\end{subfigure}
	\begin{subfigure}{0.45\textwidth}
		\centering
		\includegraphics[width=\textwidth]{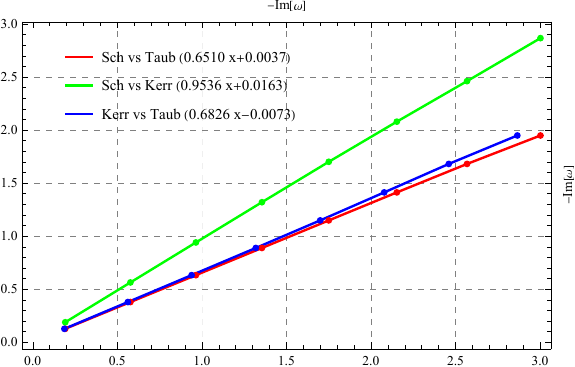}
		\caption{}
		\label{fig:scalar-massive-Im-Im-l-10}
	\end{subfigure}
	\captionsetup{width=.9\textwidth}
	\caption{
		QNFs of Schwarzschild, Kerr, and Taub-NUT BHs under a massive scalar field perturbation.
		The mass $m_{0}=0.2$, the rotation parameter $a=0.25$ for Kerr BHs, and the NUT charge $N=1$ for Taub-NUT BHs are set. Moreover, $m=0$ is set for Kerr BHs. (a) Real parts, $n=1$ and $\lambda=1, 2, \cdots, 12$. (b) Imaginary parts,  $l=10$ and $n=0, 1, \cdots, 7$.}
	\label{fig:scalar-massive-slope}
\end{figure}

\begin{figure}[!ht]
	\centering
	\begin{subfigure}{0.45\textwidth}
		\centering
		\includegraphics[width=\textwidth]{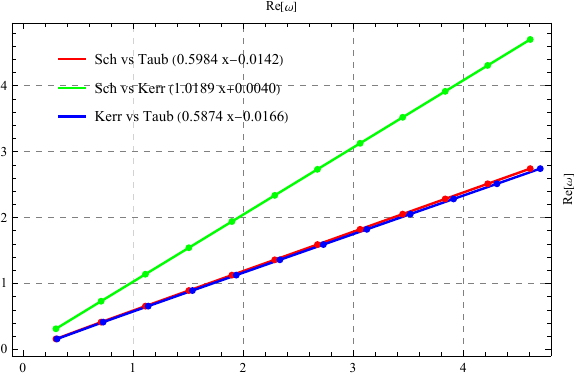}
		\caption{}
	\end{subfigure}
	\begin{subfigure}{0.45\textwidth}
		\centering
		\includegraphics[width=\textwidth]{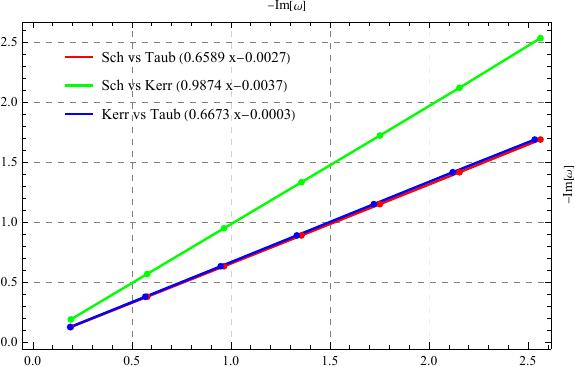}
		\caption{}
	\end{subfigure}
	\captionsetup{width=.9\textwidth}
	\caption{QNFs of Schwarzschild, Kerr, and Taub-NUT BHs under the massless spinor field-$P_{-\frac{1}{2}}$ perturbation. 
	The rotation parameter $a=0.25$ for Kerr BHs and the NUT charge $N=1$ for Taub-NUT BHs are set. Moreover, $m=0$ is set for Kerr BHs. (a) Real parts, $n=1$ and $\lambda_-=1, 2, \cdots, 12$. (b) Imaginary parts,  $l=10$ and $n=0, 1, \cdots, 6$. }
	\label{fig:dirac-massless-slope}
\end{figure}

\begin{figure}[!ht]
	\centering
	\begin{subfigure}[b]{0.45\textwidth}
		\centering
		\includegraphics[width=\textwidth]{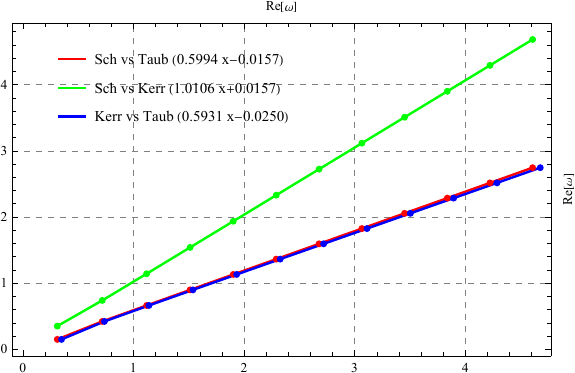}
		\caption{}
		\label{fig:Dirac-massive-Re-Re-lambda-10-s-}
	\end{subfigure}
	\begin{subfigure}[b]{0.45\textwidth}
		\centering
		\includegraphics[width=\textwidth]{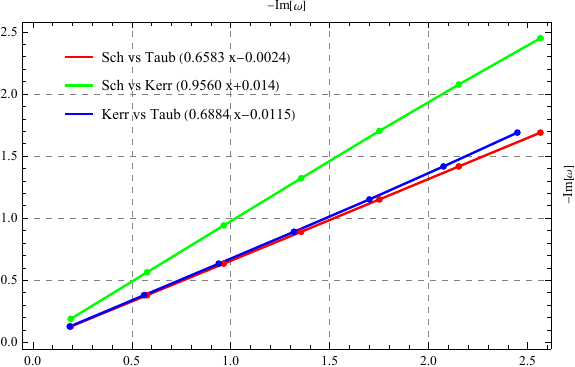}
		\caption{}
		\label{fig:Dirac-massive-Im-Im-lambda-10-s-}
	\end{subfigure}     
	\captionsetup{width=.9\textwidth}
	\caption{
		QNFs of Schwarzschild, Kerr, and Taub-NUT BHs under the massive spinor field-$P_{-\frac{1}{2}}$ perturbation. 
		The mass $m_{e}=0.1$, the rotation parameter $a=0.25$ for Kerr BHs, and the NUT charge $N=1$ for Taub-NUT BHs are set. Moreover, $m=0$ is set for Kerr BHs. (a) Real parts, $n=1$ and $\lambda_-=1, 2, \cdots, 12$. (b) Imaginary parts,  $l=10$ and $n=0, 1, \cdots, 6$.}
	\label{fig:dirac-massive-slope}
\end{figure}

In the relationships among the QNFs of the three BHs in different multipole number $l$ and overtone number $n$, we also find the linear relations of real parts between any two of the three BHs, and the linear relations of imaginary parts between any two of the three BHs,\footnote{This shows that our numerical calculations are consistent with the analytical analyses based on the light ring/QNMs correspondence in Sec.~\ref{sec:light-ring/qnms}.  
} where Fig.~\ref{fig:scalar-massless-slope} and Fig.~\ref{fig:scalar-massive-slope} correspond to massless and massive scalar field perturbations, respectively,
and Fig.~\ref{fig:dirac-massless-slope} and Fig.~\ref{fig:dirac-massive-slope} correspond to massless and massive spinor field perturbations, respectively.
The notation ``Sch vs Taub" means that the horizontal axis denotes the real (imaginary) parts of QNFs of Schwarzschild BHs, and the vertical axis stands for that of QNFs of Taub-NUT BHs. The similar meanings are taken for the others, i.e., ``Sch vs Kerr" and ``Kerr vs Taub".

Figure \ref{fig:scalar-massless-Re-Re-l-10} shows the variation of QNFs' real parts with respect to the 
separation constant\footnote{It is usual to fix the multipole number $l$ for numerical calculations. However, it is more effective if we set the separation constant $\lambda$ instead for the calculation and construction of normalizable eigenstates as mentioned in Sec.~\ref{sec:angular}. Moreover, $\lambda_-=1, 2, \cdots, 12$ for a spinor field perturbation, see also the next footnote for the explanation.}$\lambda$
that runs from $1$ to $12$, and 
Fig.~\ref{fig:scalar-massless-Im-Im-l-10} shows the variation of QNFs' imaginary parts with respect to the overtone number $n$ that runs from $0$ to $7$. Moreover, Fig.~\ref{fig:scalar-massive-slope} depicts the situation under a massive scalar field perturbation, 
Fig.~\ref{fig:dirac-massless-slope} illustrates the situation under a massless spinor field perturbation, 
and Fig.~\ref{fig:dirac-massive-slope} gives the situation under a massive spinor field perturbation. 
The points in these figures represent the numerical results and are dealt with by linear fitting, where 
the corresponding  linear functions are indicated in the figures. In addition, we note that the two components of spinor fields, $P_{\pm \frac{1}{2}}$, have the same spectra because of the symmetry.\footnote{The behaviors of $P_{\pm \frac{1}{2}}$ are akin to that of super-partners, where the spectra of  super-partners are same when a superpotential function is provided in supersymmetric quantum mechanics. For more details, refer to Ref.~\cite{Cooper:1994eh}.} 
Hence, we only focus on $P_{-\frac{1}{2}}$ in Figs.~\ref{fig:dirac-massless-slope} and \ref{fig:dirac-massive-slope}.

From Figs.~\ref{fig:scalar-massless-slope}-\ref{fig:dirac-massive-slope}, we observe that the real parts, $\Re(\omega)$, increase when the separation constant $\lambda$ or $\lambda_-$ grows, 
and the minus imaginary parts, $-\Im(\omega)$, also increase when the overtone number $n$ becomes large,
as we predict in Sec.~\ref{sec:num-qnf} in terms of the light ring/QNMs correspondence.
Specifically, when we compare Kerr BHs with Schwarzschild BHs, 
we find that the former's rotation parameter $a$ as a variable of the NJA produces the effects of increasing oscillation frequencies (the real parts of QNFs) but decreasing damping rates (the imaginary parts of QNFs); 
when we compare Taub-NUT BHs with Schwarzschild BHs, we find that the former's NUT charge $N$ as a variable of the NJA produces the effects of weakening both the oscillation frequencies and damping rates. In addition, 
the fitting results, i.e., the slopes of fitting lines are consistent with the the analytical estimations, Eqs.\ \eqref{eq:slope-ks}, \eqref{eq:slope-ns}, and \eqref{eq:slope-kn}.

\section{Conclusions}
\label{sec:conclusion}

It is an interesting topic to examine the physical reasons behind the mathematical connections established by the NJA, as mentioned in Refs.~\cite{Arkani-Hamed:2019ymq,Emond:2020lwi}. 
In this paper, we explore the  connections through a dynamical behavior of BHs, i.e., the QNFs of BHs under field perturbations. 
We find that these relationships are more than just mathematical operations and believe that the physical manifestations will be verified by future observations of GWs.

Our key finding is that the BHs linked by the NJA also exhibit a connection in their QNFs. In addition,
we notice that the rotation parameter $a$ increases the oscillation frequencies, while the NUT charge $N$ decreases them in Kerr-Taub-NUT BHs. However, both $a$ and $N$ decrease the damping rates, suggesting that the NJA has a dampening effect on wave oscillations. Furthermore, we obtain the linear relations in real parts of QNFs between any two of Schwarzschild, Kerr, and Taub-NUT BHs, and also in the imaginary parts of QNFs between any two of the three BHs, which holds even beyond the eikonal limit. 
This implies that the NJA has the ability to categorize BHs. In other words, all BHs generated with the same seed using various transformations of the NJA can be seen as belonging to a single {\em NJA class}.

The QNFs of  Schwarzschild BHs divide the NJA operation into two phases in the complex frequency plane, which is referred to as the Kerr-I and Taub-NUT (Kerr-II) phases. 
The insertion of Taub-NUT BH's QNFs into the QNF plane turns the odd-number splitting of Kerr BH's QNFs into an even-number one.
Interestingly, the QNFs exhibit several similar characteristics in these two phases, where the Kerr-I phase includes the Kerr BHs with corotating and polar orbits, and the Taub-NUT (Kerr-II) phase, contains Taub-NUT BHs and Kerr BHs with counterrotating orbits. Firstly, they vary monotonically with respect to the parameter $a$ or $N$, which is distinct from the vortex shape observed~\cite{Jing:2008an,Pan:2007gs} in RN or Einstein-Maxwell dilaton-axion BHs. 
Secondly, the damping rates vanish in the limits of $a\to M$ and 
$N\to \infty$ in the  Taub-NUT (Kerr-II) phase. 
Lastly,  the QNFs of Kerr and Taub-NUT BHs never cross the barrier established by the QNFs of Schwarzschild BHs owing to the topological protection \cite{Borde:1996df}.

Our results support the viewpoint that the NJA goes beyond a mere mathematical procedure for generating additional solutions to Einstein’s equations. In our current study, we have focused exclusively on the QNMs of a specific pair of BHs linked by the NJA. To extend our findings on the QNMs’ characteristics to a broader context, we must delve deeper into how the NJA influences QNMs. To this end, we plan to investigate whether the QNMs’ distinctive properties are also exhibited in the other BHs~\cite{Erbin:2016lzq} related through the NJA.
We think that the NJA carries profound physical implications that would be observed empirically, such as the gravitational wave detectors like LIGO, Virgo, and KAGRA~\cite{KAGRA:2021vkt}, as well as from the ongoing projects such as Taiji and TianQin~\cite{Gong:2021gvw}. Because the features of the NJA are embedded in the BHs constructed by the algorithm, we compare the QNM spectra of these BHs with the corresponding properties from gravitational waves. If our theoretical predictions coincide with the experimental data, we may conclude that the NJA provides a possible way for us to construct an acceptable BH in astrophysics from an unphysical (static) seed BH.
This serves as the primary motivation of our present work, aiming to reveal the relationships that hide behind the QNFs of BHs connected by the NJA.

This NJA dynamical phase of Schwarzschild, Kerr and Taub-NUT BHs that we study in the present paper can in fact be extended to other cases. For example, the RN, Kerr-Newman, RN-NUT BHs. The QNM curve of RN BHs intersects with that of Kerr BHs in the SKT class since RN BHs do not belong to the SKT class, but to another NJA class consisting of RN/Kerr-Newman/RN-NUT BHs. Thus, all the QNM curves are expected to have a well-splitting in the RN/Kerr-Newman/RN-NUT phase diagram, where the evidence that the QNM curves of RN and Kerr-Newman BHs do not intersect has clearly been shown in Ref.~\cite{Berti:2005eb}.
Finally, it is worth mentioning that such relationships may also be investigated through alternative means, such as the study of perturbation waveforms 
and the Zeeman splittings in the Kleinian signature~\cite{Crawley:2021auj,Crawley:2023brz,Guevara:2023wlr},  which is one of our proceeding works.

\section*{Acknowledgment}

This work was supported in part by the National Natural Science Foundation of China under Grant No. 12175108. L.C. is also supported by Yantai University under Grant No. WL22B224.

\appendix

\section{Recurrence relations and coefficients}
\label{app:rec-coef}

In this appendix, we provide a compilation of the coefficients used in the recurrence relations for scalar and spinor field perturbations, where both the massless and massive cases are considered. These coefficients have been utilized in the calculation of the QNFs in Sec.~\ref{sec:relation}.
For the details of derivations, see Ref.~\cite{Konoplya:2011qq}.

\subsection{Coefficients in recurrence relations for a massless scalar field perturbation}

The horizon is
\begin{equation}
    r_{\rm H}=\frac{1}{2} \left(\sqrt{4 N^2+1}+1\right),
\end{equation}
where we have utilized the value of $M$ as 
$1/2$ in this and subsequent formulas,

\begin{equation}
  \alpha_n=  -(n+1) (n-2 \mi r_{\rm H} \omega +1),
\end{equation}
\begin{equation}
    \beta_n =\lambda +3 n^2-10 \mi n r_{\rm H} \omega -8 r_{\rm H}^2 \omega ^2-2 \mi r_{\rm H} \omega,
\end{equation}
\begin{equation}
    \gamma_n = -\lambda -3 n^2+2 \mi n (5 r_{\rm H}+1) \omega +6 n+8 r_{\rm H}^2 \omega ^2+4 r_{\rm H} \omega ^2-12 \mi r_{\rm H} \omega -3,
\end{equation}
\begin{equation}
    \delta_n = (n-2 \mi \omega -2) (n-2 \mi r_{\rm H} \omega -2).
\end{equation}

\subsection{Coefficients in recurrence relations for a massive scalar field perturbation}

In order to write the following formulas more concise, we use $m$ instead of $m_0$ to denote the mass of massive scalar fields.

\begin{equation}
  \alpha_n  = -\mi 4 (n+1) \chi^3 (n-2 \mi r_{\rm H} \omega +1),
\end{equation}
\begin{equation}
\begin{split}
    \beta_n= 
   &-2 \chi ^2 \Big\{-2 \mi \chi  \left(\lambda +3 n^2\right)-2 \omega ^2 \left(2 n-4 \mi r_{\rm H}^2 \chi +1\right)
   +m ^2 [2 n-2 \mi r_{\rm H} (r_{\rm H} \chi +\omega )+1]\\
   &-2 (2 n+1) (2 r_{\rm H}-1) \chi ^2+4 \mi r_{\rm H} \chi  \omega  [(2 r_{\rm H}-1) \chi +3 \mi n]+4 \mi r_{\rm H} \omega ^3\Big\},
\end{split}
\end{equation}
\begin{equation}
\begin{split}
    \gamma_n= 
    &m ^4 \left[-8 n+\mi \left(4 r_{\rm H}^2 \chi +8 r_{\rm H} \omega +\chi -6 \mi\right)\right]+2 m ^2 \Big\{2 \chi  \Big[\mi \left(\lambda +3 (n-1)^2\right)+6 (n-1) r_{\rm H} \omega \\
    &-\mi r_{\rm H} (5 r_{\rm H}+2) \omega ^2\Big]+2 (2 r_{\rm H}-1) \chi ^2 (2 n-2 \mi r_{\rm H} \omega -3)+3 \omega ^2 (4 n-4 \mi r_{\rm H} \omega -3)\Big\}\\
    &+4 \omega ^2 \Big\{\chi  \Big[-\mi \left(\lambda +3 (n-1)^2\right)-6 (n-1) r_{\rm H} \omega +2 \mi r_{\rm H} (2 r_{\rm H}+1) \omega ^2\Big]\\
    &+(2 r_{\rm H}-1) \chi ^2 (-2 n+2 \mi r_{\rm H} \omega +3)+\omega ^2 (-4 n+4 \mi r_{\rm H} \omega +3)\Big\},
\end{split}
\end{equation}
\begin{equation}
\begin{split}
    \delta_n=
    &m ^4 [4 n-\mi (4 r_{\rm H} \omega +\chi -8 \mi)]+4 m ^2 \Big[\omega ^2 (-3 n+2 \mi r_{\rm H} \chi +6)-2 (n-2) r_{\rm H} \chi  \omega -\mi (n-2)^2 \chi \\
    &+3 \mi r_{\rm H} \omega ^3\Big]+4 \chi  \omega ^2 \left(2 (n-2) r_{\rm H} \omega +\mi (n-2)^2-2 \mi r_{\rm H} \omega ^2\right)+8 \omega ^4 (n-i r_{\rm H} \omega -2).
\end{split}
\end{equation}

Here we have
\begin{equation}
     \chi =\sqrt{\omega ^2-m^2}.
\end{equation}

\subsection{Coefficients in recurrence relations for a massless spinor field perturbation}

\begin{equation}
    \alpha_n=(1 + n) (3 + 2 n - 4 \mi r_{\rm H} \omega),
\end{equation}
\begin{equation}
    \beta_n = -4 n^2 - 2 \lambda^2 +  n (-4 + 16 \mi r_{\rm H} \omega) + (\mi + 4 r_{\rm H} \omega)^2,
\end{equation}
\begin{equation}
    \gamma_n =(n - 2 \mi \omega) (-1 + 2 n - 4 \mi r_{\rm H} \omega).
\end{equation}

\subsection{Coefficients in recurrence relations for a massive spinor field perturbation}

In order to write the following formulas more concise, we use $m$ instead of $m_e$ to denote the mass of massive spinor fields.

\begin{equation}
\alpha _{n}=-2 (n+1) (m r_{\rm H}-\mi \lambda ) \chi ^2 (2 n-4 \mi r_{\rm H} \omega +3)
\end{equation}
\begin{equation}
    \begin{split}
        \beta_{n}=& -4 r_{\rm H}^3 m^5+4 \mi r_{\rm H}^2 \lambda  m^4+\big[20 \omega ^2 r_{\rm H}^3+20 \mi \omega  r_{\rm H}^2-2 \left(2 \lambda ^2+2 \mi n \chi +4 \mi \omega +3\right) r_{\rm H}+2\big] m^3 \\
        & +2 \lambda  \left(2 \mi \lambda ^2-10 \mi r_{\rm H}^2 \omega ^2+\mi+2 n (4 r_{\rm H}-1) \chi +2 r_{\rm H} \omega \right) m^2 \\
        & +\big\{-16 \omega  \left(\chi ^3+\omega ^3\right) r_{\rm H}^3+4 [(-3 \mi+\omega ) \chi ^3-\omega ^3 \chi -5 \mi \omega ^3] r_{\rm H}^2+[3 \mi \chi ^3-3 \mi \omega ^2 \chi \\
        &+2 \omega ^2 \left(2 \lambda ^2+4 \mi \omega +3\right)] r_{\rm H}+4 n^2 (r_{\rm H}+1) \chi ^2-2 \omega ^2+2 n \chi ^2 [-4 \mi (2 \chi +\omega ) r_{\rm H}^2\\
        &+(7-4 \mi \omega ) r_{\rm H}-1]\big\} m -\mi \lambda  \big\{-[(4 r_{\rm H}-1) (3 \mi+4 r_{\rm H} \omega ) \chi ^3]+12 n^2 \chi ^2-\omega ^2 (3 \mi+4 r_{\rm H} \omega ) \chi \\
        & +2 n [\chi  (5-12 \mi r_{\rm H} \omega )-8 \mi r_{\rm H} \omega ^2] \chi +2 \omega ^2 \left(2 \lambda ^2-8 r_{\rm H}^2 \omega ^2-2 \mi r_{\rm H} \omega +1\right)\big\} 
    \end{split}
\end{equation}
\begin{equation}
    \begin{split}
        \gamma_{n}=& \left(-4 r_{\rm H}^3+4 r_{\rm H}^2+r_{\rm H}\right) m^5-\mi \left(4 \lambda  r_{\rm H}^2+\lambda \right) m^4+2 \big\{10 \omega ^2 r_{\rm H}^3-2 \omega  (3 \mi+7 \omega ) r_{\rm H}^2\\
        & +[-2 \lambda ^2-12 \mi (n-1) \chi +\mi \omega +2] r_{\rm H}+2 \lambda ^2+2 \mi (n-1) \chi  \big\} m^3+2 \lambda  [-2 \mi \lambda ^2\\
        &+10 \mi r_{\rm H}^2 \omega ^2-2 \mi-8 (n-1) r_{\rm H} \chi +3 \omega +2 r_{\rm H} \omega  (2 \mi \omega -5)] m^2+\big\{[-16 \omega  r_{\rm H}^3\\
        &+4 (3 \mi-4 \mi n+6 \omega ) r_{\rm H}^2+(10 \mi-4 \omega ) r_{\rm H}-\mi] \chi ^3+2 (n-1) [-4 \mi \omega  r_{\rm H}^2+(8 \mi \omega -5) r_{\rm H}\\
        &+2 n (r_{\rm H}-2)+2] \chi ^2+\omega ^2 [\mi+4 r_{\rm H} (-5 \mi+6 \mi n+\omega )] \chi -2 \omega ^2 [8 \omega ^2 r_{\rm H}^3-6 \omega  (\mi+2 \omega ) r_{\rm H}^2\\
        &+\left(-2 \lambda ^2+\mi \omega +2\right) r_{\rm H}+2 \lambda ^2]\big\} m+2 \lambda  \big\{2 \left(-4 \mi \omega  r_{\rm H}^2+r_{\rm H}+1\right) \chi ^3\\
        &+(n-1) (\mi+6 \mi n+12 r_{\rm H} \omega ) \chi ^2+\omega ^2 (n (8 r_{\rm H}+4)+r_{\rm H} (-4 \mi \omega -8)-1) \chi \\
        &+\omega ^2 \left(2 \mi \lambda ^2-4 \mi r_{\rm H} (2 r_{\rm H}+1) \omega ^2+2 \mi+(10 r_{\rm H}-3) \omega \right)\big\}
    \end{split}
\end{equation}
\begin{equation}
    \begin{split}
        \delta_{n}=& [m (r_{\rm H}-1)+i \lambda ] \big\{m^4+[-8 r_{\rm H} \omega ^2+2 \mi \omega -8 \mi r_{\rm H} \omega -4 \mi (n-2) \chi +2] m^2\\
       & +\chi ^3 (3 \mi+4 r_{\rm H} \omega )+\chi  \omega ^2 (-13 \mi+8 \mi n+4 r_{\rm H} \omega )-2 (n-2) \chi ^2 (2 n-4 \mi r_{\rm H} \omega -1)\\
       &+2 \omega ^2 \left(4 r_{\rm H} \omega ^2+i (4 r_{\rm H}-1) \omega -1\right)\big\}
    \end{split}
\end{equation}

\bibliographystyle{utphys}
\bibliography{references}
\end{document}